\documentclass[manuscript]{aastex}

\shorttitle{The star formation history and chemical evolution of SFGs}
\shortauthors{Torres-Papaqui et al.}

\begin{document}


\title{The star formation history and chemical evolution of star forming galaxies in the nearby universe.}

\author{J. P. Torres-Papaqui, R. Coziol, R. A. Ortega-Minakata \& D. M. Neri-Larios}
\affil{Departamento de Astronom\'{\i}a, Universidad de Guanajuato\\
Apartado Postal 144, 36000, Guanajuato, Gto, Mexico}
\email{papaqui@astro.ugto.mx, rcoziol@astro.ugto.mx, rene@astro.ugto.mx, daniel@astro.ugto.mx}

\begin{abstract}
We have determined the metallicity (O/H) and nitrogen abundance (N/O) of a sample of 122751 Star Forming Galaxies (SFGs) from the Data Release~7 of the Sloan Digital Sky Survey (SDSS). For all these galaxies we have also determined their morphology and obtained a comprehensive picture of their Star Formation History (SFH) using the spectral synthesis code STARLIGHT. The comparison of the chemical abundance with the SFH allows us to describe the chemical evolution of the SFGs in the nearby universe (z $\leq$ 0.25) in a manner which is consistent with the formation of their stellar populations and morphologies.

A high fraction (45\%) of the SFGs in our sample show an excess of abundance in nitrogen relative to their metallicity. We also find this excess to be accompanied by a deficiency of oxygen, which suggests that this could be the result of effective starburst winds. However, we find no difference in the mode of star formation of the nitrogen rich and nitrogen poor SFGs. Our analysis suggests they all form their stars through a succession of bursts of star formation extended over a few Gyr period. What produces the chemical differences between these galaxies seems therefore to be the intensity of the bursts: the galaxies with an excess of nitrogen are those that are presently experiencing more intense bursts, or have experienced more intense bursts in their past. We also find evidence relating the chemical evolution process to the formation of the galaxies: the galaxies with an excess of nitrogen are more massive, have more massive bulges and earlier morphologies than those showing no excess. Contrary to expectation, we find no evidence that the starburst wind efficiency decreases with the mass of the galaxies. As a possible explanation we propose that the lost of metals consistent with starburst winds took place during the formation of the galaxies, when their potential wells were still building up, and consequently were weaker than today, making starburst winds more efficient and independent of the final mass of the galaxies. In good agreement with this interpretation, we also find evidence consistent with downsizing, according to which the more massive SFGs formed before the less massive ones.

\end{abstract}


\section{Introduction}\label{sec:1}

One of the most important achievements of modern astronomy is the discovery of the process of nucleo-synthesis of chemical elements in stars. Once properly understood, this process, coupled with the concept of stellar evolution regulated by the different masses of stars, gives us a unique insight about the chemical evolution of galaxies \citep{EP78,brodie91,Zaritsky94,Coz98,Coz99,henry00,pilyugin03,pilyugin11,TP11}. For instance, it is now well accepted that oxygen and sulfur are two elements produced by massive stars ($M \ge 8 M_\odot$), while nitrogen is mostly a product of lower mass stars \citep{renzini81,McCall85,EvansDopita85,garnett90}. Consequently, due to the longer time passed on the main sequence by stars with decreasing masses, we would expect some time delay between the enrichment of oxygen and that of nitrogen in galaxies having different ages \citep{matteucci85,garnett90,molla06,richer08}. Assuming the initial mass function (IMF) does not vary between spiral galaxies, such time delay, when properly documented, may thus reveal something fundamental about how these systems formed their stars; for example, allowing to distinguish between constant star formation over the formation by a sequence of stellar bursts \citep{lehnert96,Coz99,tremonti04}.

To determine the ``normal'' or standard chemical evolution of Star Forming Galaxies (SFGs) in the nearby universe (z $\leq$ 0.25), the observation and study of a large, homogeneous and statistically representative sample is required. This is where the Sloan Digital Sky Survey (SDSS) project becomes so valuable \citep{york00,hogg01,pier03}. By applying different automatic algorithms to the enormous data bank produced by the SDSS it is now possible to retrieve one of the largest and homogenous sets of spectral line ratios necessary to estimate the basic chemical abundances, and to describe the chemical evolution process of SFGs in a manner consistent with their star formation histories and morphologies.

In recent articles \citep{tremonti04,nagao06,izotov06,yin07} data from the SDSS were already used to verify the consistency of the different methods devised in the past to determine the abundance of elements. One of the difficulties encountered in these studies is related to the rarity of the [OIII]$\lambda4363$ line. Theoretically, this line was recognized as crucial in order to determine an accurate temperature for the gas in HII regions. Unfortunately, [OIII]$\lambda4363$ can only be observed in very low metallicity SFGs, which form a minute fraction of the SDSS galaxies. This fact emphasizes the importance of developing different empirical methods like $R_{23}$ or $R_{3}$ to obtain chemical information for a significantly larger sample of galaxies \citep{Pagel79,McCall84,VC92,Thurston96,thuan10}.

Recently, some authors \citep[e.g.,][]{nagao06,yin07} advocated that we need to modify some of the empirical methods applied in previous abundance studies using only galaxies where [OIII]$\lambda4363$ was observed. These authors based their claims on the fact that they found apparent significant differences in the abundances determined when they use their new calibrations. However, such point of view is somewhat problematic, as it assumes all galaxies follow the same chemical evolutionary pattern, independently of their mass or morphology, and assumes no evolution with redshift, while these are two assumptions that need to be verified separately. Moreover, other researchers in the field \citep[e.g.,][]{tremonti04,izotov06} that also tested thoroughly the empirical methods using SDSS data, have demonstrated that in general new empirical relations show results that are in good agreement with what was found before. For these reasons, but also for comparison sake with what was done in \citet{Coz99}, we choose for our study to apply the same empirical relations that were used before, but limiting our chemical study to the two most important abundance ratios, O/H and N/O, which were shown by \citet{izotov06} to be less dependent on the method adopted to determined the gas temperature.

Another important difficulty encountered in chemical evolution studies of SFGs is related with the contamination of emission lines by absorption features produced by the underlying older stellar populations. In our research we have solved this problem by subtracting a stellar population template from each spectrum, as determined by the spectral synthesis code STARLIGHT \citep{cid05}. This method also has the advantage that through the fitted templates the star formation histories (SFHs) of the SFGs can be deduced, and other metallicity-independent parameters like the stellar velocity dispersion, that combined with the effective radius can yield an estimate of the mass of the bulge. In our study we use these new information in parallel with the morphologies which were determined independently to complete our view about the chemical evolution and formation process of SFGs in the nearby universe.

This study is organized in the following manner. In Section~\ref{sec:2} we describe how our sample of SFGs was constructed and how the data for our analysis were obtained. In Section~\ref{sec:3} we present our results for the chemical abundances, and show how they varied with the mass and morphology of the galaxies. In the same section, we also explore the relation between the chemical abundances and the SFHs, and compare our results using STARLIGHT with some relevant models from Starburst 99. In Section~\ref{sec:4} we discuss our observations and propose a new interpretation. Our main conclusion can be find in Section~\ref{sec:5}. Many of our results were verified using statistical tests, which were regrouped in Appendix A.

\section{Determination of the samples and data for analysis}\label{sec:2}

The data for our study were taken from the SDSS Data Release 7 \citep{abazajian09}. Using the STARLIGHT Virtual Observatory service\footnote{http://www.starlight.ufsc.br} we retrieved the spectroscopic data for 122751 SFGs, with emission lines having a signal to noise ratio $S/N \geq$ 3 (adjacent continuum $S/N \geq$ 10) and a redshift z $\leq$ 0.25. The spectra were corrected for Galactic dust extinction and processed through STARLIGHT spectral synthesis code \citep{cid05}, producing for each galaxy a stellar population template-corrected spectrum from which emission line fluxes were measured automatically. Note that STARLIGHT estimates and applies a correction for internal dust extinction \citep{asari07}, which means that the stellar population templates, and the star formation history deduced from these templates, as well as the template-subtracted emission spectra are free of this effect.

In SDSS, the fibers have a fixed aperture and are centered on the nucleus of the galaxies. Comparing the physical projection of this aperture on the sky with the efficient radius as determined by \citet{simard11}, we found that, except at very low redshifts, the spectra always cover the same physical regions comparable with the sizes of the bulges of the galaxies. This implies that our stellar population study and chemical analysis are mostly concerned with processes affecting the bulges of the galaxies.

\subsection{Separation between SFGs and AGNs}\label{sec:2.1}

\begin{figure}
\epsscale{1.0} \plotone{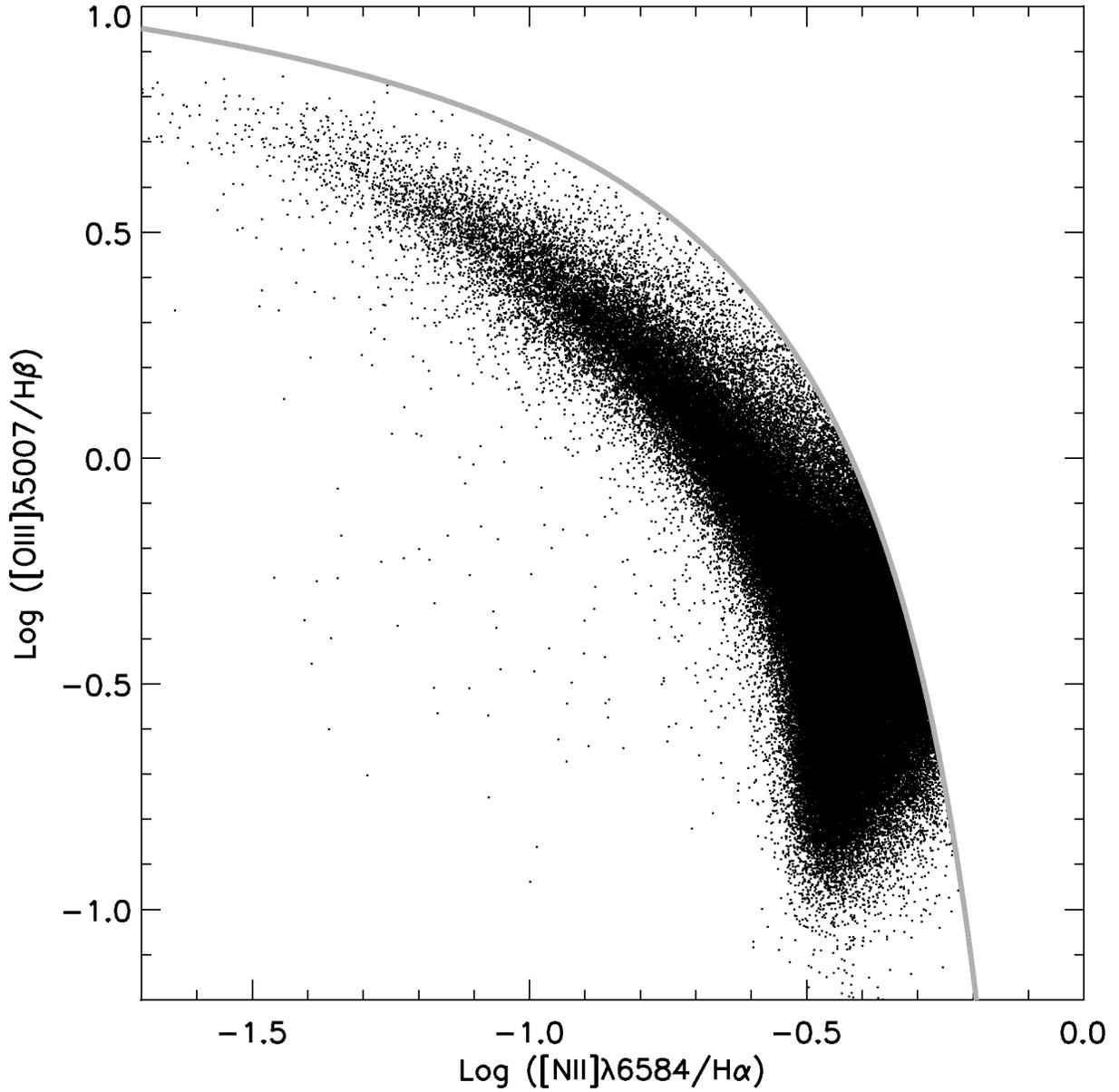} 
\caption{Diagnostic diagram for the 122751 SDSS galaxies classified as SFG by applying the empirical separation criterion (solid grey curve) proposed by \citet{kauffmann03}. According to this criterion, none of these galaxies should show an AGN contribution.}
\label{fig:1}
\end{figure}

The SFG classification presented in Figure~\ref{fig:1} is based on one diagnostic diagram \citep{baldwin81,veilleux87}, where we have applied the separation criterion between AGNs and SFGs proposed by \citet{kauffmann03}. According to the standard interpretation of stellar ionized HII regions \citep{McCall85,EvansDopita85,Coz96}, the SFGs trace a sequence of increasing metallicity as the emission ratio [OIII]$\lambda$5007/H$\beta$ decreases, which is produced by the cooling effect of oxygen. Similarly, the abundance of nitrogen is also expected to grow in the same direction, as the ratio [NII]$\lambda$6584/H$\alpha$ increases \citep{Thurston96,vanZee98}.

\begin{figure}
\epsscale{1.0} \plotone{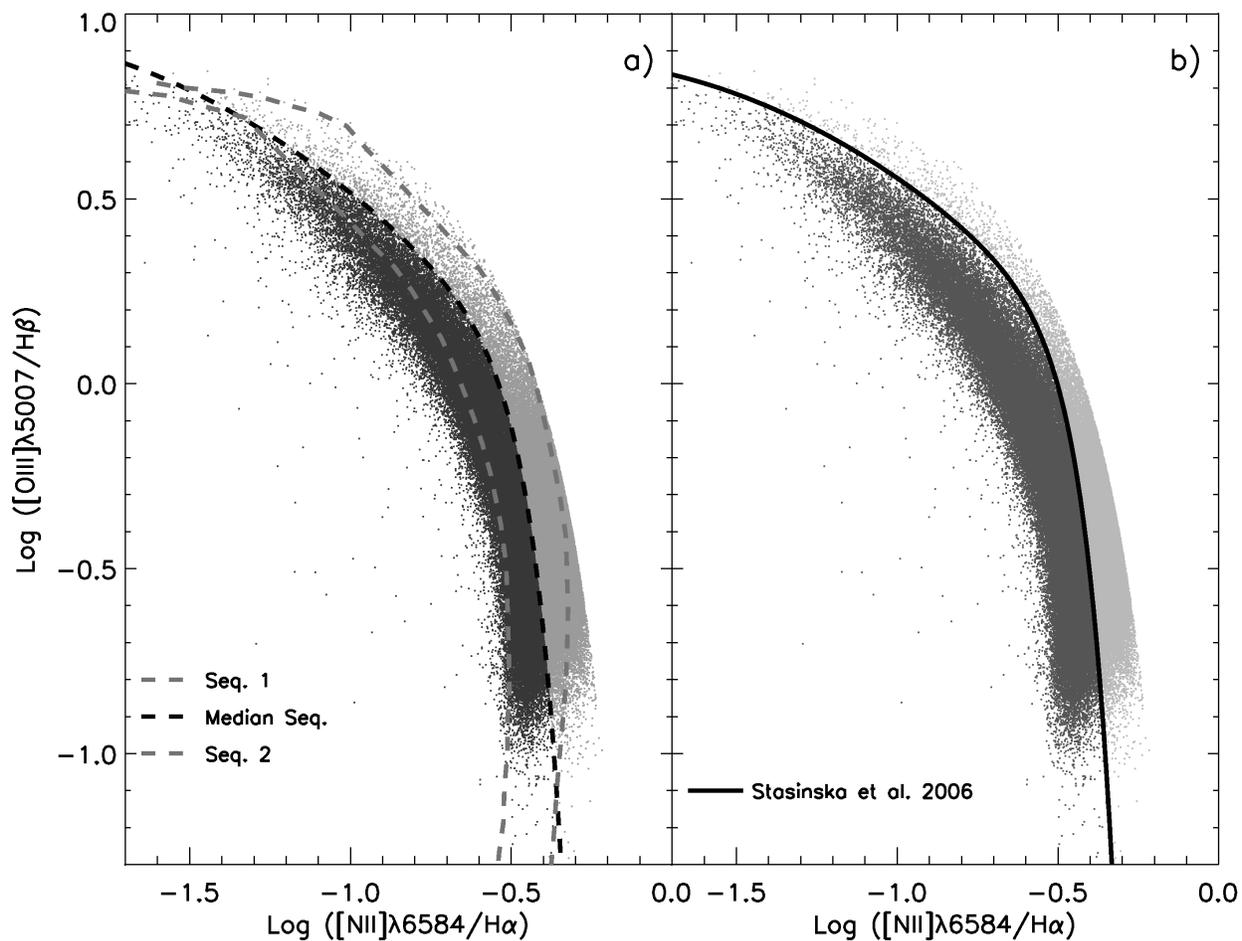} 
\caption{Same as Figure~\ref{fig:1} using different separation criteria: a) according to \citet{Coz99} models of HII regions, where Seq.~1 is a standard HII region abundance model, and Seq.~2 a similar model including an excess in nitrogen abundance by 0.2 dex--the median of these models (bold dashed curve) separates nitrogen poor from nitrogen rich SFGs; b) according to an alternate criterion to separate SFGs from AGNs, as proposed by \citet{stasinska06}.}
\label{fig:2}
\end{figure}

Consistent with the standard interpretation, \citet{Coz99} have determined that some SFGs, especially those experiencing a starburst in their nuclear regions, present an excess of nitrogen abundance compared to the expected chemical evolution sequence traced by HII regions in late-type spirals. This is illustrated in Figure~\ref{fig:2}a, where we show the results of two ionization models as produced by \citet{Coz99}: for the same metallicity, galaxies following the Sequence~2 (Seq.~2) have a nitrogen abundance about 0.2 dex higher than galaxies following the Sequence~1 (Seq.~1). Figure~\ref{fig:2} reveals that most of the SFGs in our sample have a data point that falls between these two models. Using the median of these two models, we have thus separated our SFG sample in two groups (see Figure~\ref{fig:2}a). The SFGs that are nitrogen poor are identified as SFG~1, and represent 55.2\% of the sample, while those that are nitrogen rich are identified as SFG~2, and represent 44.8\%.

In Figure~\ref{fig:2}a we notice that the path of Seq.~2 seems to form an upper limit rather than a median as was determined by \citet{Coz99}. The reason for this difference is because in 1999 the concept of transition objects (TOs), which are  defined roughly as galaxies where both an AGN and intense star forming activity are present, was not applied systematically in classification studies, and the limit between SFGs and AGNs was then located at higher values of the ratio [NII]$\lambda$6584/H$\alpha$. Although \citet{Coz99} in their analysis presented evidence against the presence of an AGN in their sample, this possibility remains in this much larger and differently selected sample. Therefore, the SFG~2 galaxies may show an excess of excitation due to an AGN instead of an excess of nitrogen abundance. Interestingly, in Figure~\ref{fig:2}b we can see that the median of the two abundance models defined in (a) almost coincides with a different separation criterion between SFGs and AGNs which was previously proposed by \citet{stasinska06}. If we use this distinction we could separate our sample into two different new groups, according to which 64.8\% of the galaxies are pure SFGs (SFG~3), and the rest (35.2\%) are AGNs (SFG~4).

For our sample of SFGs we have therefore identified two different interpretations for the increase of nitrogen emission in the standard diagnostic diagram. The first one suggests some SFGs show an excess in nitrogen abundance, the other suggests an excess in emission due to a supplementary source of ionizing photons, consistent with a central AGN. In order to discriminate which of these two hypothesis is the most probable, we here present the results of two tests. The first test, which was devised by \citet{Coz99}, consists in comparing the ratio [NII]$\lambda$6584/H$\alpha$ with the ratio [SII]$\lambda\lambda$6717,6731/H$\alpha$. The principle of this test is that if there is an excess of excitation, this excess should appear in both line ratios at the same time.

\begin{figure}
\epsscale{1.0} \plotone{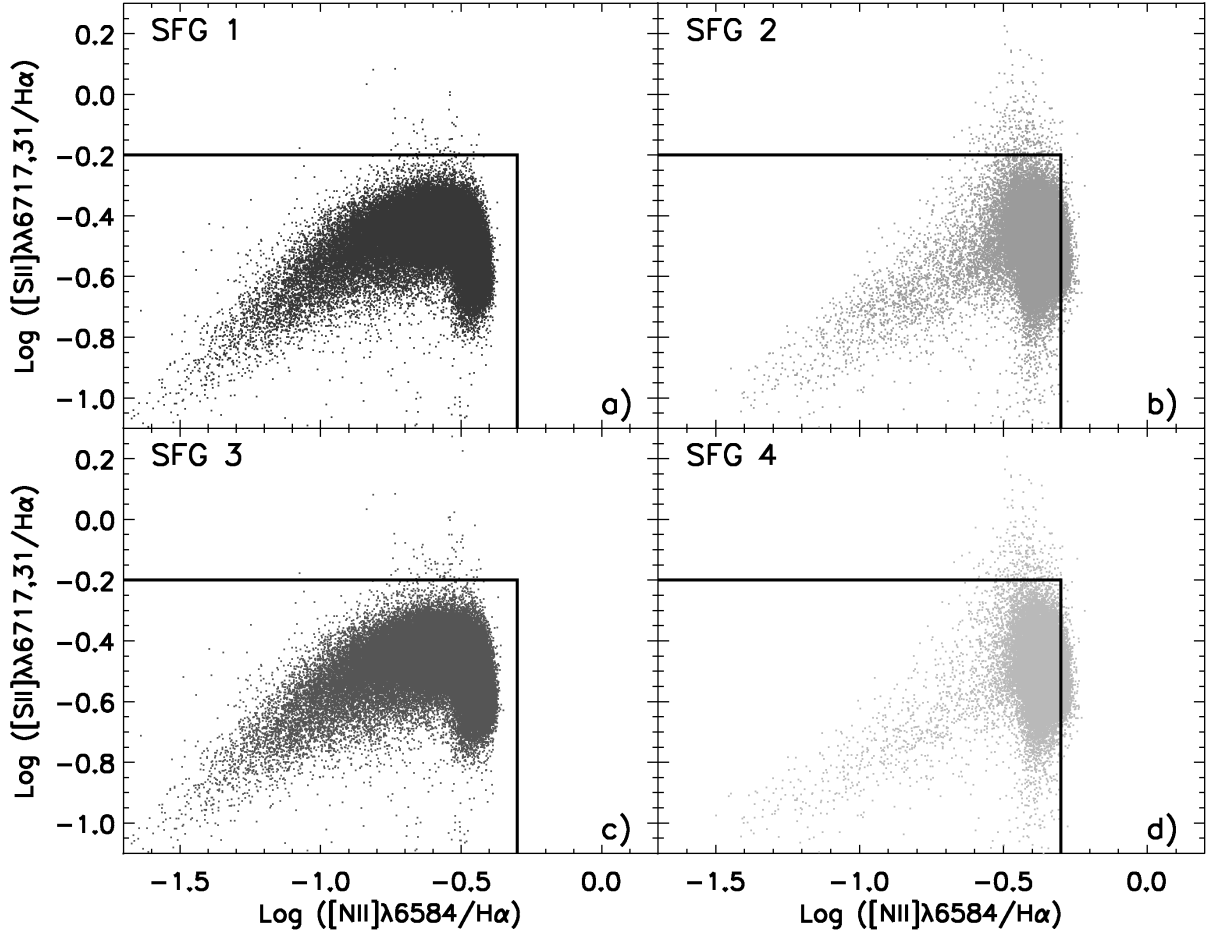} 
\caption{Excitation test, to check for the presence of AGNs, as proposed by \citet{Coz99}. The lower left boxes contain galaxies with normal level of excitation (due to massive stars), while galaxies showing an excess of excitation in both lines fall out of the limits of these boxes. The four subsamples correspond to nitrogen poor SFGs (SFG 1) vs. nitrogen rich SFGs (SFG 2), according to \citet{Coz99}, and pure SFGs (SFG 3) vs. AGNs (SFG 4) according to Stasinska et al. (2006).}
\label{fig:3}
\end{figure}

We show the results of applying the first test to our SFG samples in Figure~\ref{fig:3}. The effect for an excess of excitation is not observed in the SFG~1~(a), nor than in the SFG~3~(c). Instead, we observe that at the same time the emission of nitrogen increases, the sulfur emission decreases. The same trend is also observed in Figure~\ref{fig:3}(b) for the SFG~2, and in (d) for the SFG~4, although now most of the galaxies are well into the decrement phase of sulfur. Only an insignificant number of galaxies in these two samples are consistent with an excess of excitation in both lines. From this first test we conclude that, in general, we see no evidence for an excess in excitation that could have been produced by an AGN.

\begin{figure}
\epsscale{1.0} \plotone{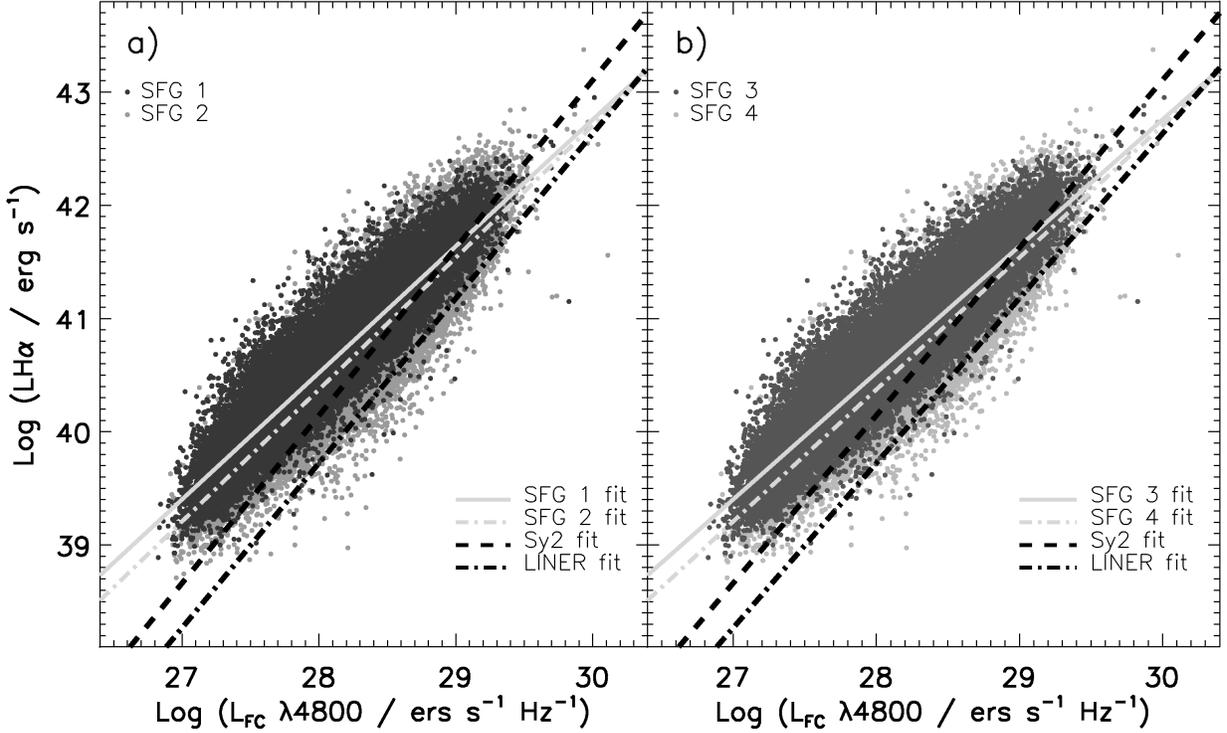} 
\caption{Second test for the presence of AGN based on the relations between the logarithm of the luminosity of the continuum ($\rm{L_{FC}}$) and the logarithm of the luminosity of the ionized gas (L$_{\rm{H}\alpha}$): a) results for the SFG~1 and SFG~2 samples; b) results for SFG~3 and SFG~4 samples. The significantly steeper linear relations for the Seyfert~2 and LINERs were determined by \citet{TP12}. For numerical values see Table~\ref{table1}.} 
\label{fig:4}
\end{figure}

As a second test, we present in Figure~\ref{fig:4}a for the SFG~1 and SFG~2 and in Figure~\ref{fig:4}b for the SFG~3 and SFG~4 the graph of the H$\alpha$ luminosity as a function of the luminosity of the continuum at 4800 \AA. The continuum fluxes were measured in the raw spectra, that is, before they were treated by STARLIGHT, and were only corrected for Galactic extinction. The principle of this test is based on a suggestion made by \citet{osterbrock89}, according to who the linear relation (in Log) between the two luminosities must be different for galaxies excited by AGNs and galaxies excited by stars, because of the different powers laws in these objects relating the continuum with the ionizing flux. In Figure~\ref{fig:4} we observe that the linear relations for our four SFG groups have almost the same slopes. These linear relations have extremely high factors of correlation (see Table~\ref{table1}). For comparison we have traced the relation found for two sample of SDSS AGNs classified as Seyfert~2 and LINERs \citep{TP12}. The slopes of the linear relations for the SFGs are significantly shallower than for those measured in AGNs. From this second test we conclude, once again, that there is no evidence for the presence an AGN in our sample of SFGs.

Taking into account the high level of consistency of the above two tests, we conclude that in general we see no evidence for AGNs in our sample of SFGs, and we find consequently no reason to change our classification based on the separation criterion proposed by \citet{kauffmann03}. For the remaining of our analysis we will keep only the distinction between nitrogen poor (SFG~1) and nitrogen rich (SFG~2) SFGs.

\subsection{Determination of abundances and physical parameters}\label{sec:2.2}

Once the possibility of an AGN as a source of ionization for the gas is eliminated, standard methods for HII regions can be applied to determine the chemical abundances of the SFGs. For the metallicity, O/H, we used the method described in \citet{VC92}, which yields values with a typical uncertainty of the order of $\pm0.2$ dex \citep{EP84}. For the nitrogen abundance, N/O, we utilized the method described in \citet{Thurston96}, where the temperature of the gas, T$_{II}$, is first estimated using the ratio $([\rm{OII}]\lambda\lambda3727,3729 + [\rm{OIII}]\lambda\lambda4959,5007)/\rm{H}\beta$, which is then combined with the ratio $[\rm{NII}]\lambda\lambda6548,6584/[\rm{OII}]\lambda\lambda3727,3729$ to estimate the ratio N/O. According to \citet{McCall85}, the dispersions in the distributions of these two line ratios are due to a variation in metallicity (O/H) and relative nitrogen abundance (N/O), which justify the abundance determination method. A correction for ionization may become important only at low metallicity, 12 + log(O/H) $<$ 8.2 \citep{izotov06}, which is not the case for the galaxies in our sample. When necessary, all our abundances and results from models are given relative to the new Sun metallicity, as determined by \citet{Asplund04}.

From the STARLIGHT template we retrieved the mean age of the stellar populations, from which we can deduce the SFH of the galaxy. The SFH describes how the star formation rate (SFR) in a galaxy has varied over its lifetime. Since the stellar templates are corrected for internal extinction the SFR are also free of this effect. To produce the SFR we worked with a smoothed version of the population vector obtained from the output of STARLIGHT \citep{asari07}, where at each time, $t_i$, the SFR($t_i$) is composed of a mixture of stars taken from six different metallicity groups. As discussed by \citet{asari07} in their section 5, any value in the range 25 Myr bin for the stellar library yields a strong correlation with the luminosity in H$\alpha$, which is used to measure the current SFR. However, this bin in age is not critical in STARLIGHT synthesis code, because values in the whole 10-100 Myr yield correlations of similar strength (this excludes WR stars which have much shorter ages, but there is no trace of such stars in our spectra). This range of ages is fully consistent with what is used in other synthesis codes, like for example Starburst~99. We therefore expect our results using STARLIGHT to be fully comparable and consistent with Starburst~99 models (see Section~3.3).

Also from the STARLIGHT templates, we have retrieved the stellar velocity dispersion of the stars, $\sigma_{\star}$. After correcting for the instrumental resolution \citep[following][]{greene06}, the $\sigma_{\star}$ are used in combination with the effective radii determined by \citet{simard11} to estimate the masses of the bulges (applying the virial theorem). Note that within the limits in redshift of our study the spatial projection of the aperture of the SDSS fiber varies from 1 kpc to 7 kpc, but at the same time the effective radii of the galaxies also increase, which implies that the fraction of the galaxy covered by the fiber at any redshift is roughly the same. Because the projected apertures are comparable with the effective radii of the galaxies where the kinematic of the bulge is dominant, no correction for the effect of rotation was applied to $\sigma_{\star}$. 

As an independent parameter (that is, not depending on STARLIGHT), we have classified all the galaxies in our sample adopting a morphological index T, varying on a scale from -5 to 10 ($-5=$\ E, $10 =$\ Irr). The Hubble morphological types are presented in Table~\ref{table2}, together with the correspondence adopted between the de Vaucouleur's morphological index \citep{Vaucouleurs1991} and our own. We have determined the morphology using the correlations between photometric colors, the inverse concentration index and the morphological types, as was found by \citet{shimasaku01} and \citet{fukugita07}. The photometric colors are $u-g$, $g-r$, $r-i$, and $i-z$, which are defined in the photometric system of the SDSS\footnote{http://casjobs.sdss.org}. The inverse concentration corresponds to $R_{50}(r)/R_{90}(r)$, that compares the Petrosian radii \citep{petrosian76} containing 50\% and 90\% of the total flux in the $r$ band. A K-correction was applied to the Petrosian magnitudes using the code developed by \citet{BR07}.

A last independent parameter useful for our analysis is the absolute magnitude in B, which was retrieved from the SDSS Data Release 7 \citep{abazajian09}. Being a good indicator of star formation activity over relatively long periods of time, from 0.5 up to 6 Gyr \citep{Coz96}, the luminosities in B can also serve as a proxy for the total mass of the galaxies. Considering the low limit in redshift of our study, a simple cosmology was applied, adopting an Hubble constant H$_0 = 75$ km s$^{-1}$ Mpc$^{-1}$.

\section{Results}\label{sec:3}

\subsection{Variation of abundances, and relation with mass and morphology}\label{sec:3.1}

\begin{figure}
\epsscale{0.9} \plotone{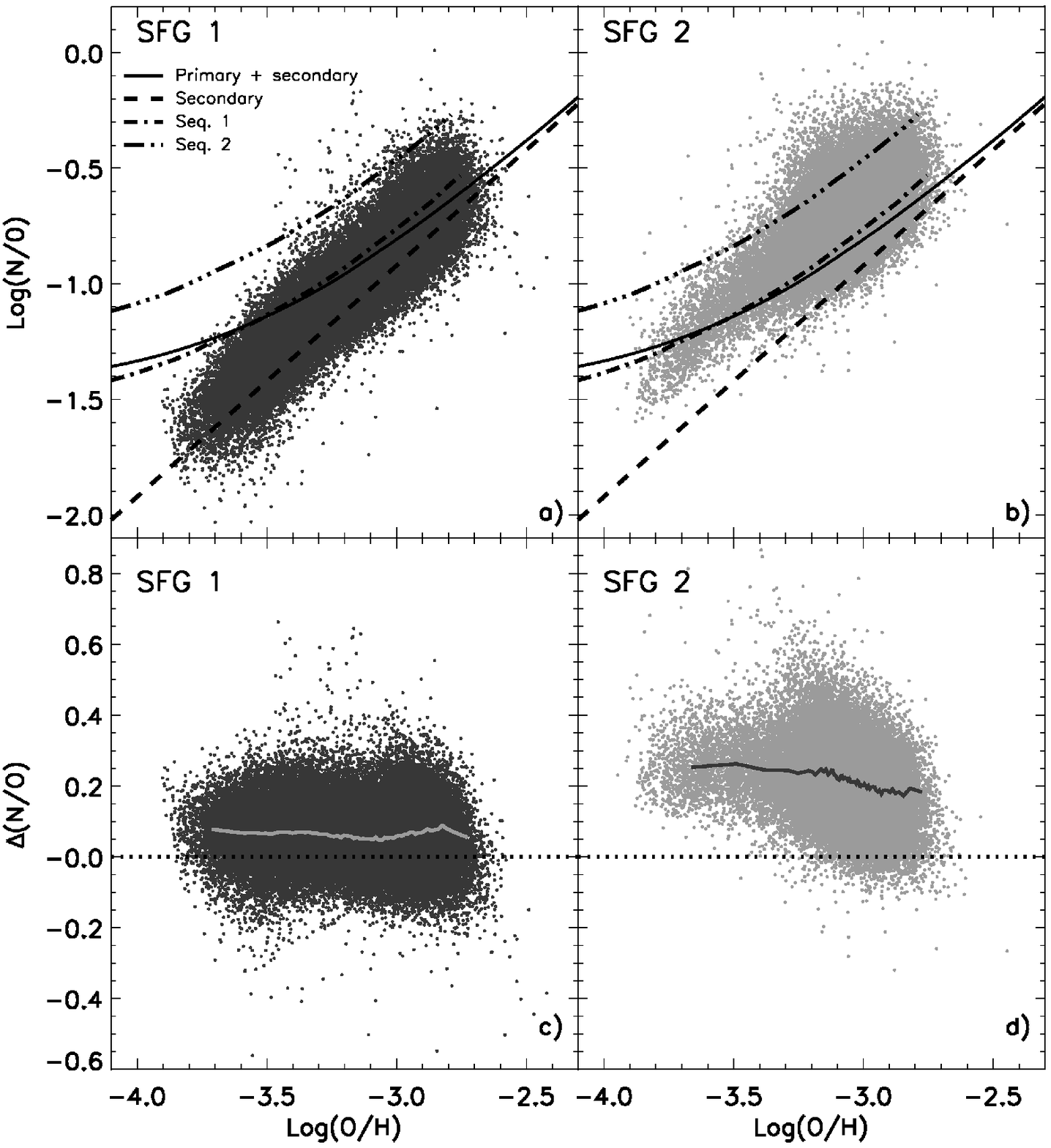} 
\caption{Nitrogen abundance as a function of metallicity: in a) the SFGs poor in nitrogen (SFG~1), and in b) the SFGs rich in nitrogen (SFG~2). Also shown are the closed-box chemical evolution models proposed by \citet{VE93}: the $secondary$ model (dashed line) and the $primary + secondary$ model (continuous curve). The two other curves trace the sequences obtained in \citet{Coz99}. In c) and d) we show the same results relative to the secondary relation for nitrogen enrichment, with their respective medians.}
\label{fig:5}
\end{figure}

In Figure~\ref{fig:5} we trace the abundance of nitrogen as a function of the metallicity, distinguishing between nitrogen poor (SFG~1) and nitrogen rich (SFG~2) SFGs. Overplotted on this figure we show the two abundance sequence models obtained by \citet{Coz99}. Also shown are the relations for the different closed-box chemical evolutionary models for the production of nitrogen as proposed by \citet{VE93}: the $secondary$ and the $primary + secondary$ models. In general, the SFG~1 seem to follow the secondary enrichment model. This is obviously not the case for the SFG~2, however, which for the same value in $\log$ (O/H) show an excess of nitrogen, on average of 0.15 dex,
compared to the SFG~1.

In Figure~\ref{fig:5}b we also note a remarkable feature: the chemical enrichment process for the SFG~2 is not continuous--we observe an abrupt jump by 0.3 dex in nitrogen at a metallicity $\log$ (O/H) = $-3.4$ (almost the solar metallicity). The reason why we note this particularity is because exactly the same observation was done before by \citet{Coz99} for their comparatively small sample of about 100 UV bright SBNGs. The fact that exactly the same feature appears in our significantly larger sample and generally selected sample (no other criterion other than the SFG classification was applied) is quite remarkable. It suggests that this feature must be a characteristic of SFGs (it affects ~45\% of all the galaxies in our sample), and not a peculiar trait of starburst galaxies. This phenomenon must be related consequently to some common mechanism or event occurring during the chemical evolution of all these galaxies.

In Figure~\ref{fig:5}a, although the nitrogen distribution for the SFG~1 fits the secondary relation, we may detect a slight trend toward relatively high nitrogen abundance also in this sample. This is better perceived in Figure~\ref{fig:5}c where we have plotted over the data the medians of the difference in nitrogen abundance relative to the secondary relation. In \citet{Coz99} the UV bright starburst galaxies were showing a trend toward a decrease in the excess of nitrogen at high metallicity. The same trend may also be visible in Figure~\ref{fig:5}d for the SFG~2, although, from the medians, we conclude that the decrement is much weaker than what was observed in \citet{Coz99}.

\begin{figure}
\epsscale{1.0} \plotone{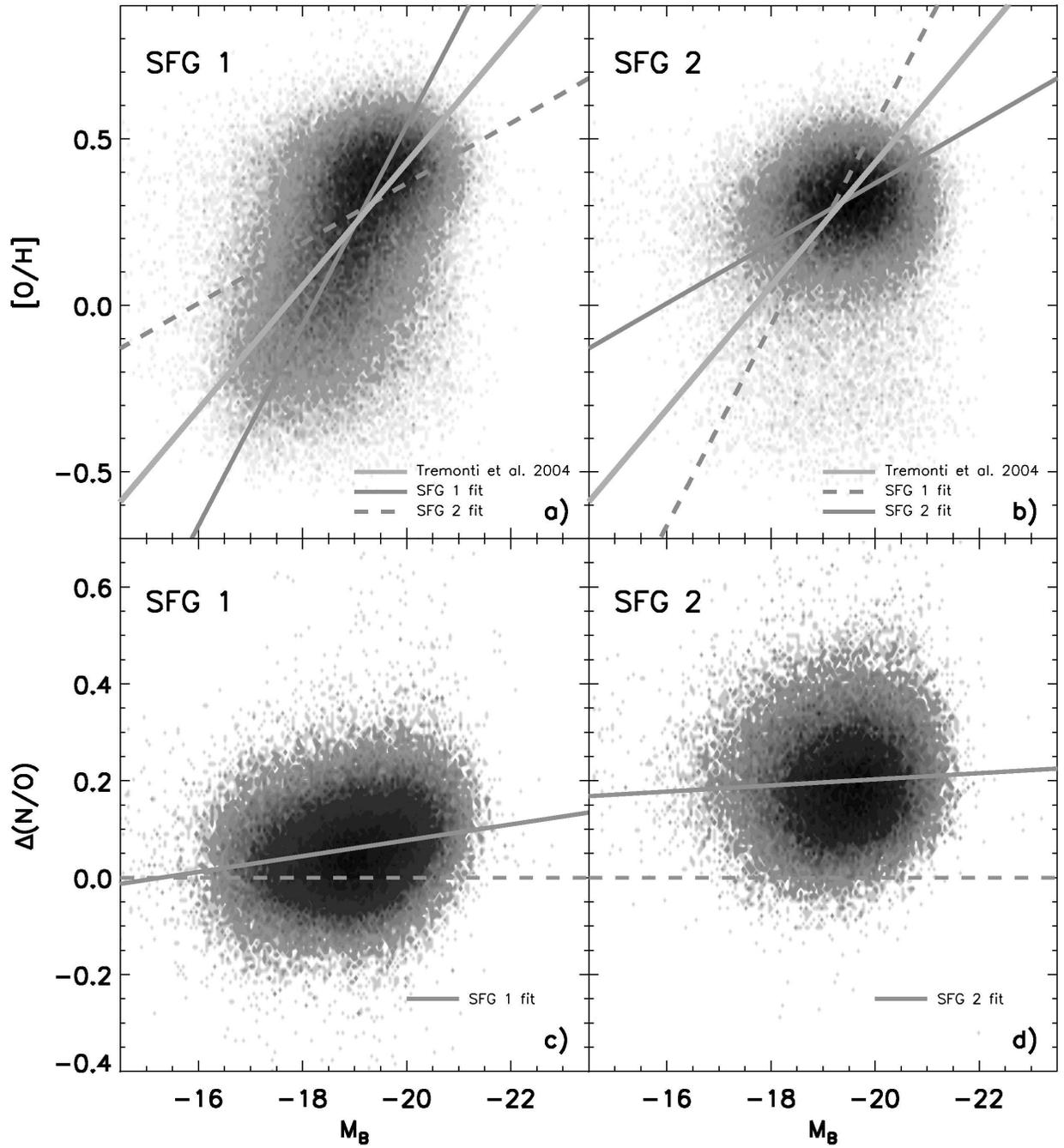} 
\caption{Mass-metallicity and mass-nitrogen excess relations for the SFG~1 and SFG~2. Overplotted on the data in a) and b) we compare our linear correlations with the relation obtained by \citet{tremonti04}. Results for the correlation tests are presented in Table~\ref{table3}.}
\label{fig:6}
\end{figure}

\begin{figure}
\epsscale{1.0} \plotone{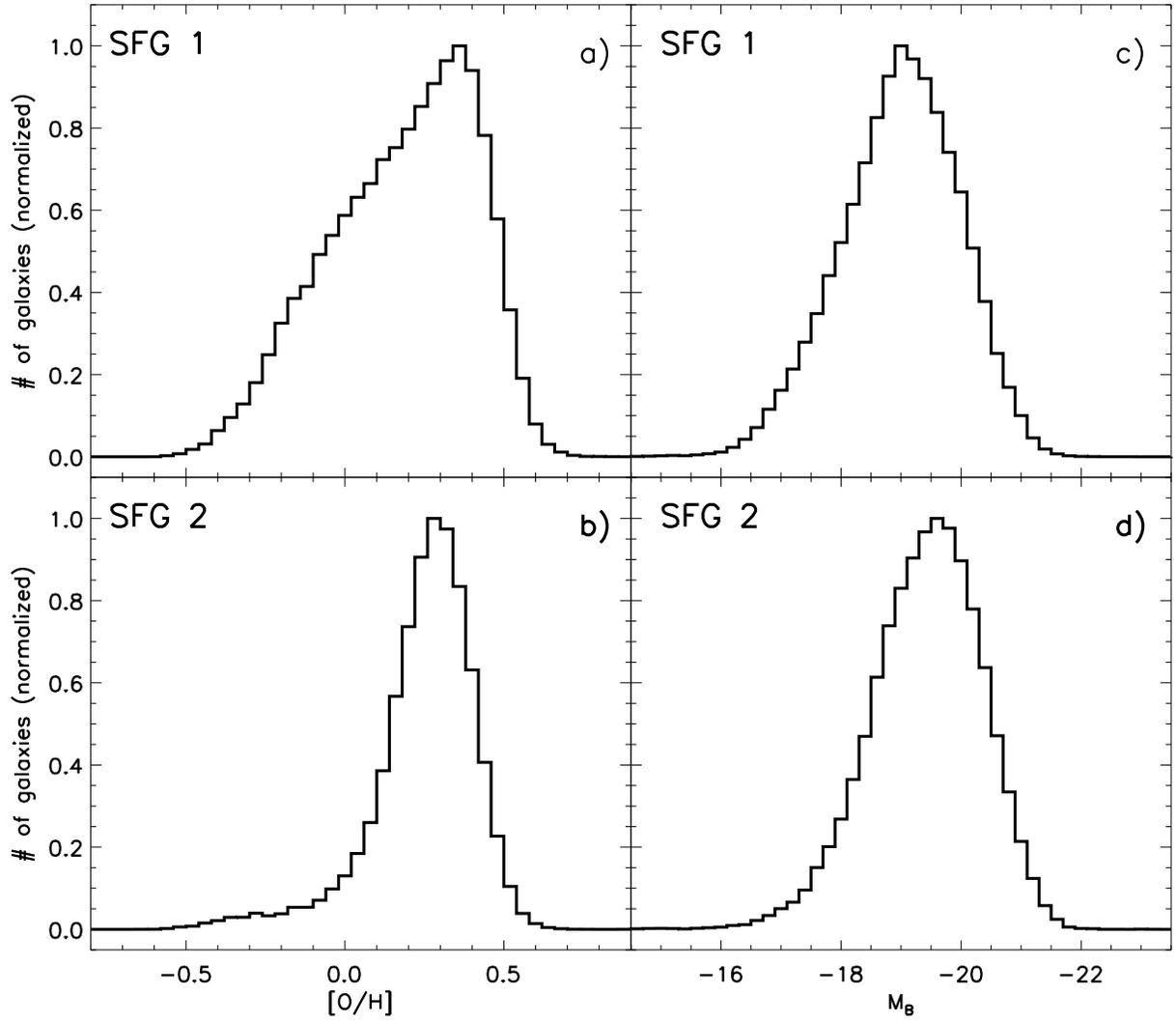} 
\caption{Histograms for the metallicities and absolute magnitudes in the SFG~1 and SFG~2.}
\label{fig:7}
\end{figure}

In Figure~\ref{fig:6} we trace the metallicity and nitrogen offset from the secondary relation as a function of the absolute magnitude in B. In Figure~\ref{fig:6}a and b, we have included the luminosity-metallicity relation as determined by \citet{tremonti04} for their sample of 53400 SDSS SFGs (correcting for the Hubble constant used in our study). Our samples taken as a whole (adding the SFG~1 with the SFG~2) are in good agreement with this relation. However, when we introduce the distinction between nitrogen rich and nitrogen poor SFGs we obtain a different perspective. The metallicity of the SFG~1 increases with the mass more rapidly than the relation found by \citet{tremonti04}, while it increases less rapidly in the SFG~2. In fact, and as the correlation tests in Table~\ref{table3} are confirming, the metallicity of the SFG~2 stays almost constant. Therefore, it is like the SFG~2 become deficient in oxygen as the mass increases. This is verified in Figure~\ref{fig:7}, which shows the different histograms for the metallicities and absolute magnitudes. Although the SFG~2 are on average more massive (more luminous in B) than the SFG~1, their peak in metallicity lays at a lower value than in the SFG~1. This confirms that the SFG~2 are relatively deficient in oxygen as compared to the SFG~1. Note that the same phenomenon was also observed by \citet{Coz97} for the same sample of UV bright starburst galaxies that were shown later to be nitrogen rich \citep{Coz99}. So it seems now that the two phenomena must be somehow connected.

In Figure~\ref{fig:6}c we find that the excess of nitrogen in the SFG~1 is slightly increasing with the mass (although in Table~\ref{table3} we have a 36-38\% percent chance that the correlation is spurious). On the other hand, in Figure~\ref{fig:6}d for the SFG~2 the excess stays constant independently of the mass. This is confirmed by the correlation tests (Table~\ref{table3}). It seems therefore like the chemical abundances in the SFG~2 reach some sort of physical limit connected with the masses of the galaxies.

In Figure~\ref{fig:8} we compare the results obtained for the morphologies, the mean stellar population ages, the masses of the bulges, and absolute magnitudes in B, as measured in the SFG~1 and SFG~2 samples. The box-whisker diagrams show that the SFG~2 have earlier-types (Figure~\ref{fig:8}a) and older stellar populations than the SFG~1 (Figure~\ref{fig:8}b). The masses of the bulges in Figure~\ref{fig:8}c are also increasing in the SFG~2, which is consistent with the earlier morphologies. The difference in absolute magnitude in B (Figure~\ref{fig:6}d) also implies higher masses for the SFG~2 as compared to the SFG~1. All the above differences were confirmed at a level of confidence of 95\% by statistical tests (see Figure~\ref{fig:22} in appendix~A). All these variations and differences point toward a connection between the chemical evolution process and the formation process of the SFGs.

\subsection{Analysis of SFH using STARLIGHT}\label{sec:3.2}

\begin{figure}
\epsscale{1.0} \plotone{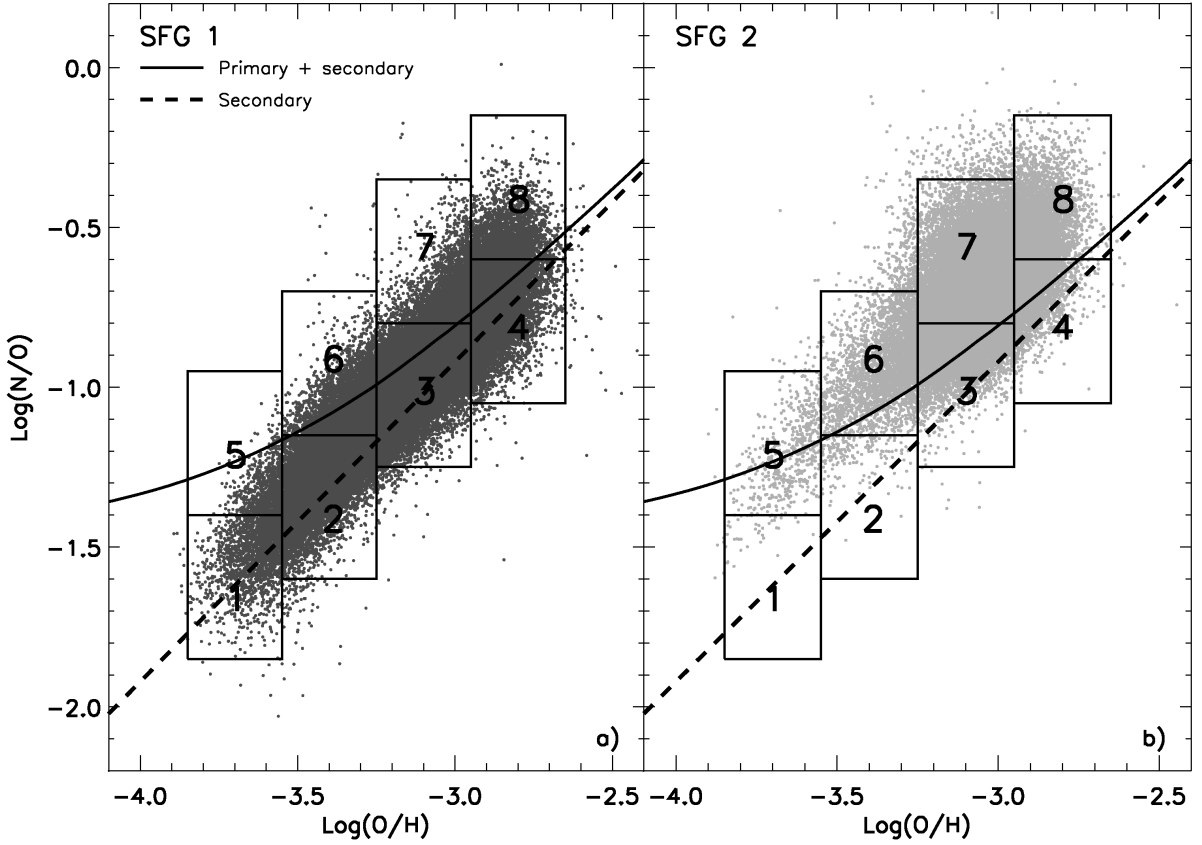} 
\caption{Variation of abundances, where we have separated the whole sample in 4 subsamples with increasing metallicity. In each subsample we further distinguish between low (subsamples 1, 2, 3, and 4) and high (subsamples 5, 6, 7, and 8) nitrogen abundances.}
\label{fig:9}
\end{figure}

\begin{figure}
\epsscale{0.7} \plotone{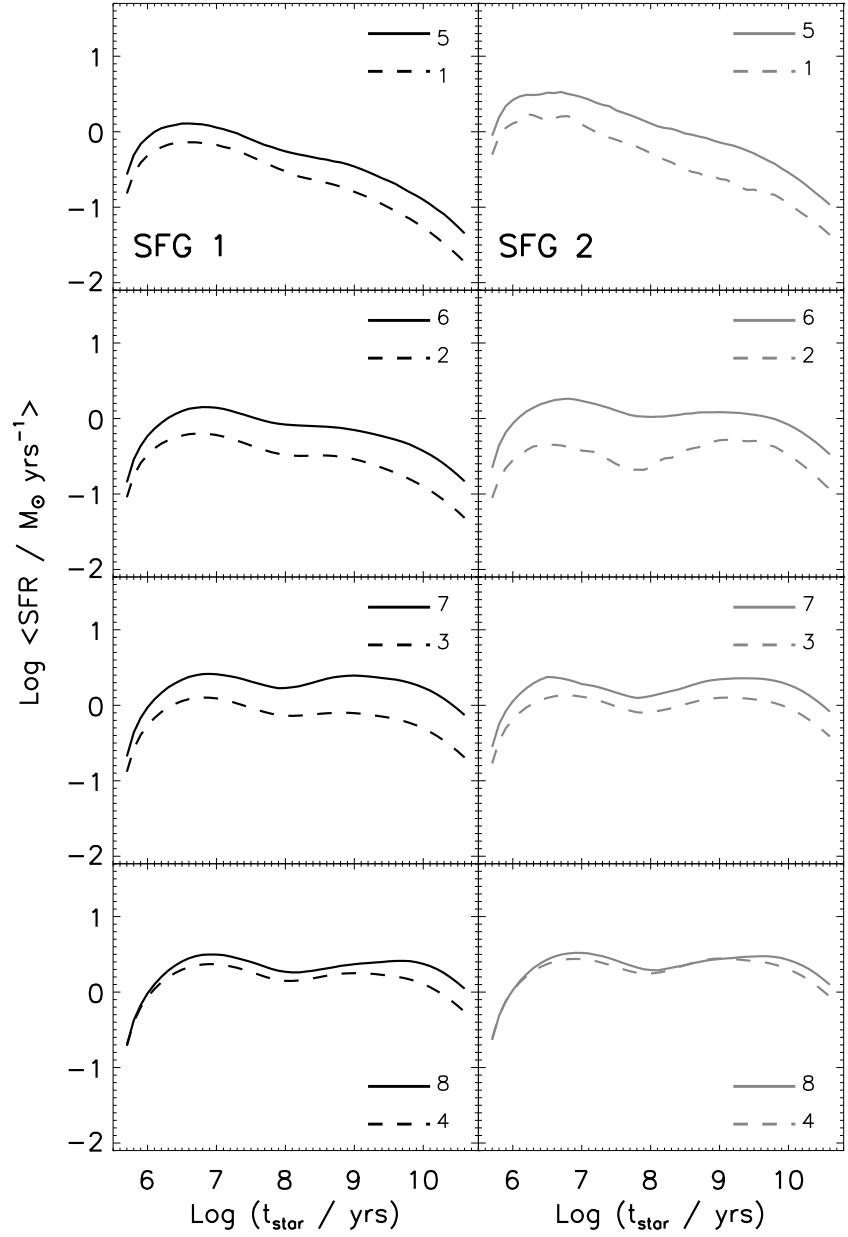}
 \caption{Variation of the star formation history (SFH) in the different subsamples defined in Figure~\ref{fig:9}.}
\label{fig:10}
\end{figure}

Through the fitting templates using STARLIGHT we have deduced the history of star formation (SFH) for all the galaxies in our sample. In order to investigate how the chemical evolution in the SFGs correlates with their SFHs, we have separated the diagram of abundances in Figure~\ref{fig:9} in 4 subsamples with increasing metallicity. In each metallicity bin we further distinguish between low and high nitrogen abundance, obtaining in total  8 different subsamples. The corresponding SFH for these 8 subsamples are shown in Figure~\ref{fig:10}. The median physical characteristics of the galaxies in each subsample are reported in Table~\ref{table4} for the SFG~1 and Table~\ref{table5} for the SFG~2.

The SFH describes how the star formation rate (SFR) of a galaxy varied over its lifetime. In particular, the SFH allows to distinguish between galaxies that have experienced more or less constant star formation rates (CSFR), as is expected for late-type spirals, and galaxies in a present starburst phase (SBP). Galaxies with CSFRs have a flat SFH (e.g., subsamples 4 and 8 in Figure~\ref{fig:10}) while galaxies in SBP, with recent SFR higher than in its past, have a negative SFH slope (e.g., subsamples 1 and 5 in Figure~\ref{fig:10}). Consequently, a galaxy with a positive SFH slope would have formed most of its stars in the past, as it is expected for early-type spirals and ellipticals (no case is observed in our sample). The SFH also yields information about the intensity of star formation: for any interval of time the SFR corresponds to the median, which implies the higher the median the more intense the star formation at this time; consequently, the higher the intensity of star formation and the higher the mass of stars formed, which is equal to the surface below the curve.

For the SFG~1 sample in Figure~\ref{fig:10} (left panels) we observe that almost all the galaxies with low nitrogen abundance (in subsamples 1, 2 and possibly 3) show some evidence of recent bursts of star formation. For the SFG~1 that are rich in nitrogen, we observe a transition to constant star formation starting with subsamples~7 (galaxies in subsamples 5 and 6 are still in bursts). In the SFG~2 (right panels in Figure~\ref{fig:10}) the sharp increase in nitrogen abundance seems connected with intense (more intense than in the SFG~1) bursts in star formation (subsamples 1, 5 and 6). We also note that the transition to constant star formation appears earlier than in the SFG~1, in subsample~2 instead of 3. In general, we observe that galaxies having higher intensity bursts or higher SFR at any time, end up with higher nitrogen abundance (comparing the continuous curves with the dashed ones). However, the difference seems to decrease at high metallicity (e.g., subsamples 4 and 8 in both the SFG~1 and SFG~2 samples).

The box-whisker plots for the SFG~1 in Figure~\ref{fig:11} and for the SFG~2 in Figure~\ref{fig:12} are consistent with our previous analysis based on the abundances (see also Table~\ref{table3} and Table~\ref{table4}). In both samples, we observe a change toward earlier-type morphology and older stellar populations as the metallicity and nitrogen increases. We also observe an increase in bulge mass and total mass. In general, the more massive and early-type SFGs with low metallicity are those that are experiencing more intense bursts, and at high metallicity the more massive and early-type SFGs are also those that have experienced higher level of star formation in the past. All these differences were confirmed by statistical tests in appendix~A (see Figure~\ref{fig:23} and Figure~\ref{fig:24}).

\subsection{Comparison with Starburst 99 models}\label{sec:3.3}

The results of our analysis based on STARLIGHT were compared using the model for star forming galaxies Starburst 99 \citep{leitherer99}\footnote{http://www.stsci.edu/science/starburst99/docs/default.htm}. We used some of the original 1999 dataset, considering only instantaneous or constant star formation scenarios, and have run a few more models using the WINDOWS version of the software. The main goal of this comparison is to verified if we can obtain Starburst 99 models that are consistent with our STARLIGHT analysis, in particular, by isolating the galaxies that show recent starbursts. Note that we did not performed an exhaustive study using Starburst 99, because this would have fall well beyond the limits of the present study. However, we have found that just a few simple modifications to the parameters of the original dataset were sufficient to yield results in good agreement with STARLIGHT.

\begin{figure}
\epsscale{1.0} \plotone{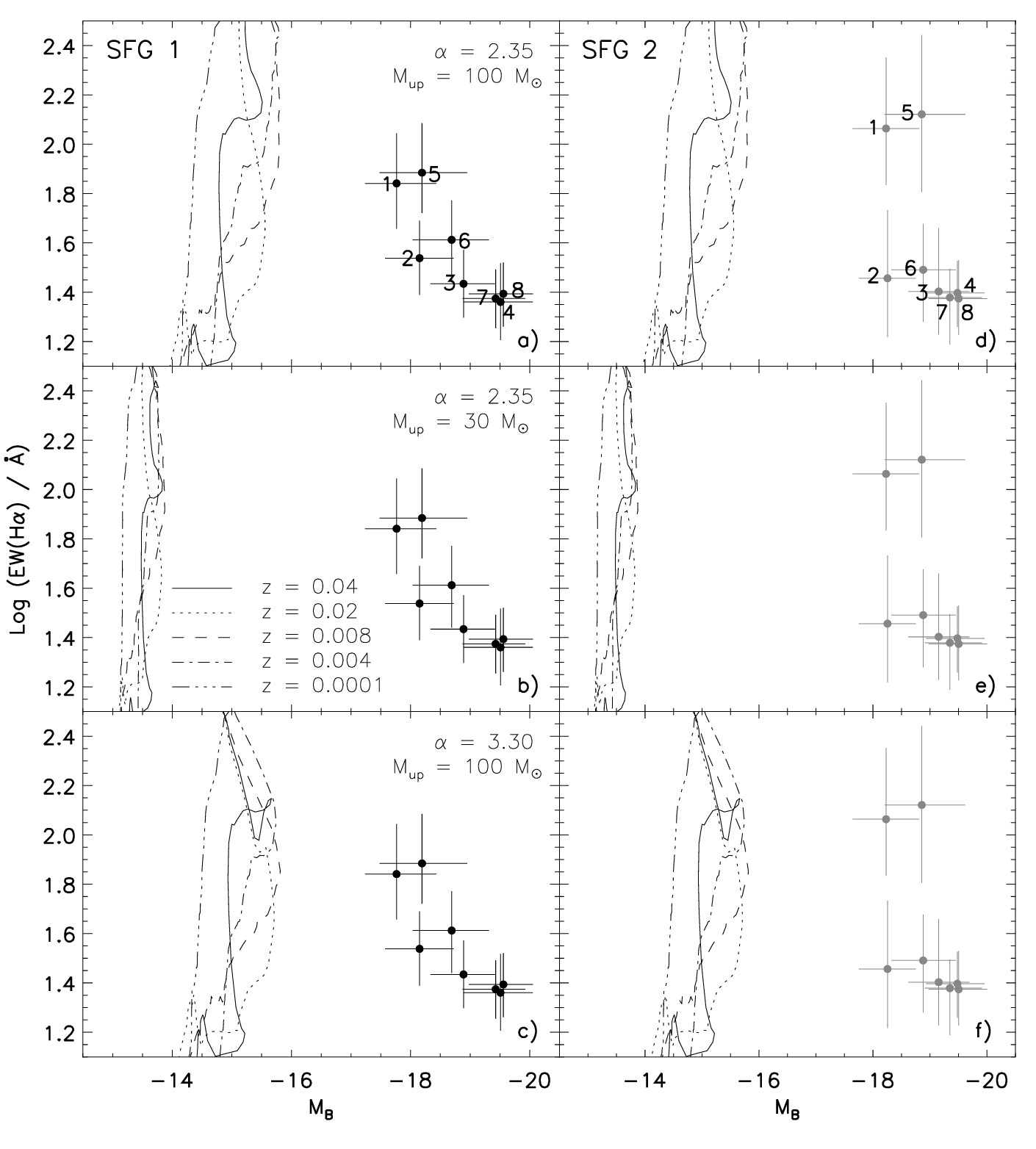} 
\caption{Results of Starburst 99 models assuming instantaneous bursts with different IMF powers and upper mass limits. The data correspond to the medians and percentiles as measured in the subsamples in metallicity as defined in Figure~\ref{fig:9} (identified by their numbers).}
\label{fig:13}
\end{figure}

The only two parameters produced by Starburst 99 that can be compare with our results from STARLIGHT are the H$\alpha$ equivalent width and absolute magnitude in B. The first parameter is a good indicator for star formation over short timescales, for this is the ratio of the flux in emission with the flux in the continuum. As we already mentioned, the second parameter is a good indicator for star formation over longer timescales (0.5 to 6 Gyr), and a reasonable proxy for the total mass of galaxies. In Figure~\ref{fig:13}, we show the predictions of instantaneous bursts models, having different powers of the IMF ($\alpha$) and upper mass limit (M$_{up}$); the Salpeter IMF corresponds to $\alpha = 2.35$, and only two mass limits were tested, 30 and 100 M$_\odot$. The data presented are the medians and percentiles as measured in the galaxies found in the abundance subsamples defined in Figure~\ref{fig:9}. It is obvious that the absolute magnitudes of the SFGs are too high to be reproduced by an instantaneous burst scenario. This result clearly suggests that in these galaxies star formation happened over longer time scales.

\begin{figure}
\epsscale{1.0} \plotone{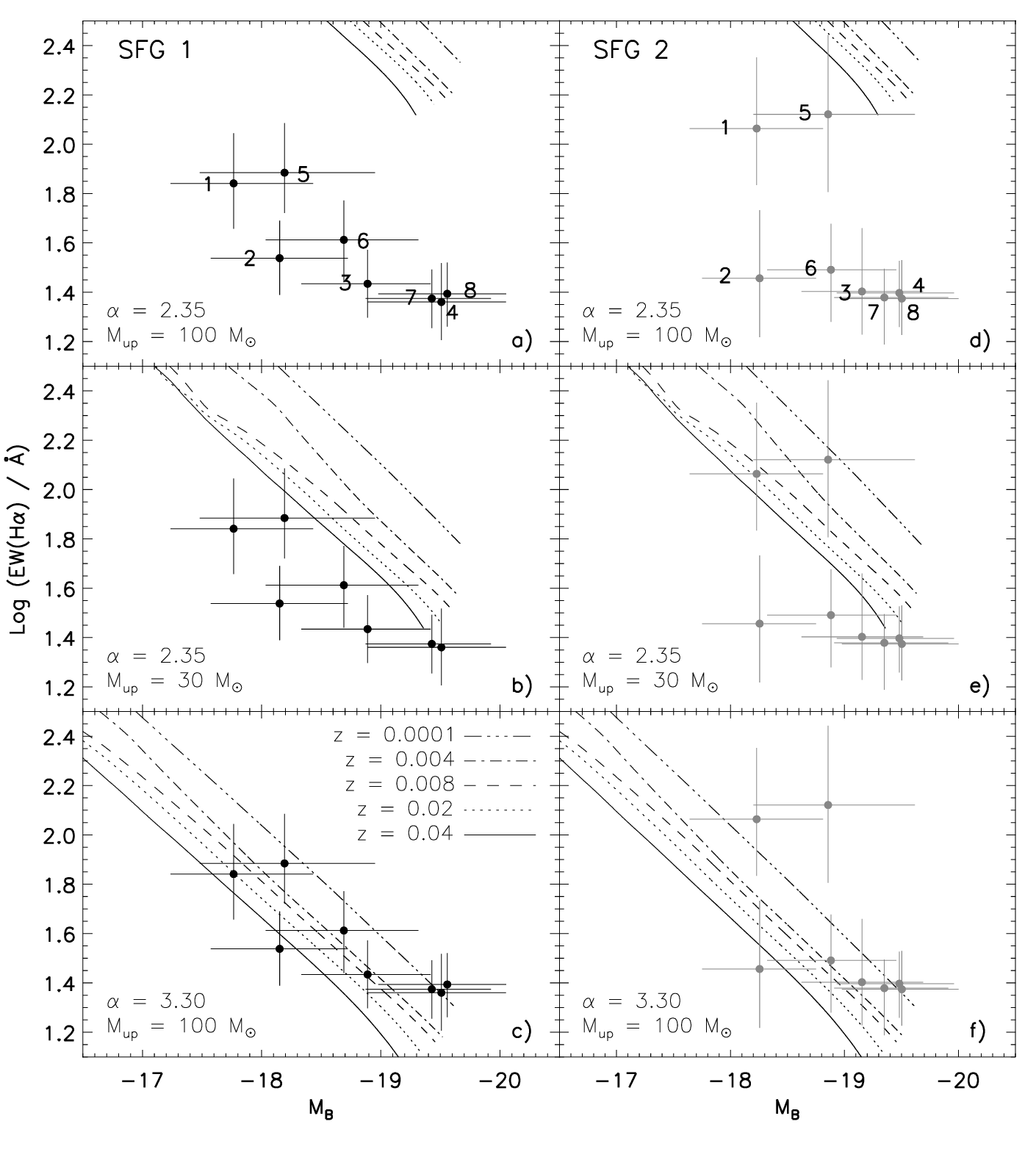} 
\caption{Results of Starburst 99 models assuming continuous star formation with different IMF powers and upper mass limits. The data correspond to the medians and percentiles as measured in the subsamples in metallicity as defined in Figure~\ref{fig:9} (identified by their numbers).}
\label{fig:14}
\end{figure}

\begin{figure}
\epsscale{1.0} \plotone{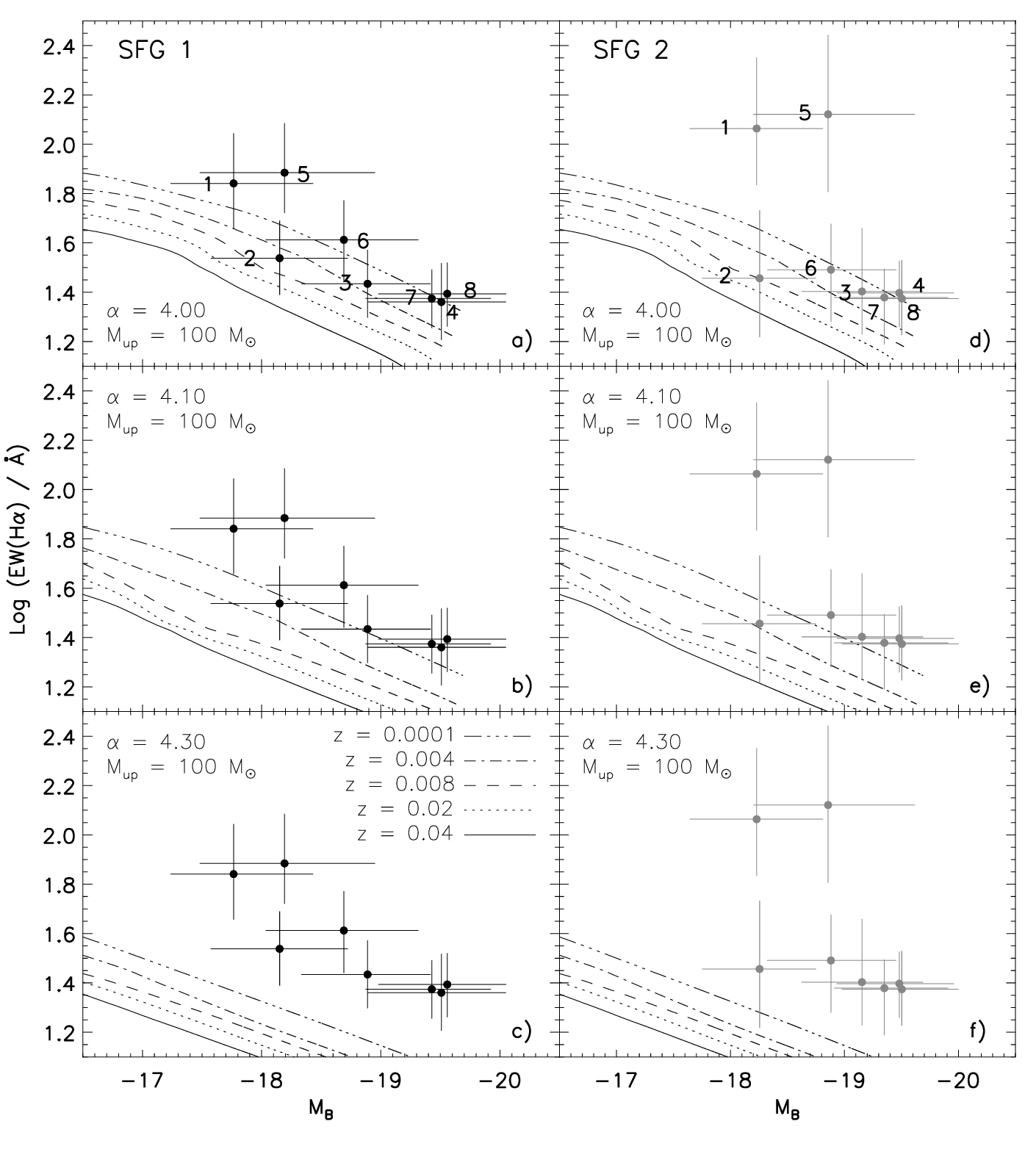} 
\caption{Same as in Figure~\ref{fig:14} for different IMF powers and upper mass limits.}
 \label{fig:15}
\end{figure}

The next series of models in Figure~\ref{fig:14} are for constant star formation. The first good fits are obtained with a steeper IMF than Salpeter, $\alpha = 3.30$, and upper mass limit M$_{up}=100\ $M$_\odot$. But, then we loose most of the SFGs that STARLIGHT was identifying as starbursts. Based on this series of models it becomes obvious in which direction we need to modify the parameters to obtain fits that are consistent with STARLIGHT analysis. The IMF has to become steeper still, implying that there are less massive stars and more intermediate mass stars ionizing the gas in these galaxies (more B than O stars). But then to compensate, the total amount of ionizing photons produced has also to be higher, which justify the upper mass limit M$_{up}=100\ $M$_\odot$. Our two best fits, which yield almost a one to one relation with the STARLIGHT results are presented in Figure~\ref{fig:15}. They have M$_{up}=100\ $M$_\odot$ and an IMF varying from $\alpha = 4.00$ (a and d) to $\alpha = 4.10$, (b and e). On the other hand none of these models reproduce the trend in metallicity we observed.

Based our comparison, we do not claim we have determined that the IMF in the SFGs is much steeper than Salpeter or that the upper mass limit is 100 M$_\odot$. What we suggest, rather, is that the results of Starburst 99 that are most consistent with STARLIGHT implies recent bursts of star formation happening over a background of constant star formation. But, considering that constant star formation in galaxies can be easily reproduced by a sequence of instantaneous bursts \citep{leitherer99}, therefore, it seems that the best scenario that the models of Starburst 99, consistent with STARLIGHT, imposes on the data, can be, for both SFGs, a sequence of bursts of star formation over a relatively long period of time. Similar results were obtained before in starburst studies like \citet{goldader97} and \citet{CDD01}. There is also now mounting evidence that typical starburst galaxies, like M 82, have experienced more than one burst spread both in time and space \citep{deGrijs01,Smith01,ForsterSchreiber03a,ForsterSchreiber03b,Smith06,Mayya06,Strickland09}. That this scenario applies in general to starburst galaxies was already verified by \citet{CDD01}. These last authors even confirmed the dominant presence of B type stars in the ionizing regions of these galaxies, which is a natural consequence of a sequence of bursts over a few Gyr period \citep{Coz96}.

From our comparison with Starburst 99 models we conclude that our STARLIGHT analysis (except for the metallicity) yields results that are consistent in all the SFGs with a sequence of bursts of star formation over a long period of time (of Gyr scale). 

\section{Discussion}\label{sec:4}

In \citet{Coz99}, the authors suggested that the discontinuous pattern observed in the diagram of nitrogen vs. metallicity was the result of a special mode of star formation in starburst galaxies. Assuming these galaxies experienced a succession of bursts over a few Gyr period, decreasing in intensity with time, and assuming a time delay between the production of oxygen by massive O and B stars and the production of nitrogen by intermediate mass stars \citep{garnett90}, then the higher the intensity of the first bursts, the higher the increase of nitrogen compared to the oxygen.

Although our present observations seem to be in good agreement with the above starburst scenario, our new analysis proposes a slightly different interpretation. First, our sample of SFGs is more general than the sample defined in \citet{Coz99}. The only criterion applied in the selection of the galaxies is that they must not show any evidence of ionization from an AGN. Second, our analysis of the SFHs produced by STARLIGHT suggests that there is no dichotomy in the star forming mode of the SFGs, that is, all the SFGs follow the same star formation pattern, consistent with a sequence of bursts of star formation over a period of a few Gyr. The comparison of our STARLIGHT analysis with Starburst 99 models is extremely compelling on this point. What our new analysis suggests, therefore, is that the nitrogen excess is not an anomaly related to some special galaxies like starbursts, but is more likely a typical characteristic of SFGs in the nearby universe. This also implies that the phenomenon that creates the excess of nitrogen must be extremely common in star forming galaxies, although possibly more obvious in starburst galaxies.

Consistent with the above interpretation, we believe that the common phenomenon responsible for the depletion of oxygen and relative increase in nitrogen as observed in the SFGs could be starburst winds \citep{Heckman90,lehnert96}. In particular, this mechanism is suspected to be the main phenomenon responsible for increasing the abundance of metals in the intergalactic medium \citep[see][and references therein]{Veilleux05}. Now, it is easy to adapt the model of a sequence of bursts as proposed by \citet{Coz99} including as a major component the effect of starburst winds. We assume that a starburst wind in a region is effective (that is, capable of ejecting metals out of the region) only for a short period of time, of the order of a few $10^7$ yrs (as is confirmed using Starburst 99). Consequently the main effect of this wind would be to preferentially deplete the region of the chemical components produced by massive stars (mostly oxygen, neon, iron, but also sulfur, argon and calcium; see Strickland \& Heckman 2009, and references therein for a full discussion of starburst wind loads). Then, when intermediate stars begin producing nitrogen, a few $10^8$ yrs after the beginning of the burst, the intensity of the wind would already be to low to have a significant impact on the abundance. Moreover, within the sequence of bursts model we do not expect the next generation of bursts to affect the nitrogen abundance in the region, simply because these bursts will happen in different regions, considering that in such a brief interval of time, a few $10^8$, the conditions in a region that suffered a burst would not be favorable for another one. However, these conditions could easily become favorable in another nearby region, consistent with the propagation of star formation hypothesis; after $10^8$ yrs active star formation regions would have propagated a few kpc from the first burst region \citep[see for example][and references therein]{Nomura01}.

Can we verify the starburst wind hypothesis directly with our observations? To answer this question we refer to the detailed calculation made by \citet{Strickland09}, according to who the difference between the nitrogen rich and nitrogen poor SFGs should depend on the intensity of the bursts. In their model, a starburst wind is considered to be effective above an intensity of 0.04 M$_{\odot}\ \rm{yrs}^{-1} \rm{kpc}^{-2}$ \citep{lehnert96,Heckman03}. Using STARLIGHT we have estimated the intensity of the bursts (see Table \ref{table6}) for the galaxies in both samples showing the higher recent SFR (subsamples 1 and 5). In Figure~\ref{fig:16}, we show the corresponding box-whisker plots. We see that the intensity of the bursts in the galaxies with the most obvious excess of nitrogen (the SFG~2 in subsample~5) are significantly higher than in galaxies without an excess (SFG~1). Also shown in this figure is the threshold calculated by \citet{Strickland09}. The median and average of the SFG~2 are well above this limit. Note that we should not expect a clear cut distinction, considering that these bursts are only the most recent of a series. However, statistically the difference is significative, and in good agreement with the starburst wind hypothesis.

In Table~\ref{table6} we also give the statistics for the absolute magnitudes in B. The SFG~2 are more luminous in B than the SFG~1. This is consistent with other observations in starburst galaxies suggesting that in general the intensity of a starburst increases with the mass of the galaxy \citep{goldader97,CDD01}. Also consistent with this difference of characteristic, the STARLIGHT analysis in Figure~\ref{fig:10} reveals that in the past the SFG~2 have also experienced more intense star forming episodes than the SFG~1. All these differences were found to be significant using statistical tests (see Figure~\ref{fig:25} in appendix A). In general, our results are quantitatively consistent with the predictions made by the starburst wind model.

On the other hand, one observation seems somewhat counterintuitive in regard to the starburst wind hypothesis. Many authors have discussed and concluded that the higher the mass of a galaxy and the more difficult it should be for this system to loose metals through starburst winds \citep[e.g.][]{tremonti04}. Yet, our observations show no evidence of such trend. In Figure~\ref{fig:6}d for the SFG~2 the excess of nitrogen stays constant independently of the mass, and in Figure~\ref{fig:6}c we may even detect an increase of nitrogen with the mass in the SFG~1. On this matter, our analysis of STARLIGHT results in parallel with our determination of the morphologies and mass of the bulges is quite revealing. It allows us to connect the excess of nitrogen with massive galaxies, having developed massive bulges and earlier type morphologies. At the same time it shows that the star formation in the past of these galaxies was more intense than in those without the nitrogen excess. There is consequently a very strong relation with the formation process of the galaxies. This may suggest that the abundance differences observed between the SFG~2 and SFG~1 took place mostly during the formation process of these galaxies. This could have happened when the gravitational potential wells of these galaxies were less massive, because still growing in size, making the intense starburst winds slightly more efficient and independent of the final, present mass.

Consistent with the formation hypothesis, it was recently reported by \citet{pilyugin11} that the SFGs are showing evidence for downsizing, which states that massive galaxies formed at higher redshifts than smaller mass galaxies. If we can find such evidence in our sample of galaxies, consistent with the scenario we proposed for the excess of nitrogen, then we would have established another link between the chemical evolution and the formation process of these galaxies.

To test the downsizing hypothesis in our sample we have separated the SFG~1 and SFG~2 samples in 9 redshift subsamples increasing by 0.025. The different bin ranges are given in Table~\ref{table7} for the SFG~1 and Table~\ref{table8} for the SFG~2, together with the number of galaxies in each redshift subsamples. Note that our limit in redshift is lower than in \citet{pilyugin11}, z$=0.25$ instead of z$=0.4$. In Figure~\ref{fig:17} we show the box-whisker plots for the variations of the properties of the SFG~1 as a function of the redshift. The same graph is shown for the SFG~2 in Figure~\ref{fig:18}. In the SFG~1 we see an increase in metallicity and nitrogen excess at higher redshift. This is not observed however in the SFG~2 sample, where the metallicity and nitrogen excess stay almost constant (see Figure~\ref{fig:26} and Figure~\ref{fig:27} in appendix A for confirmation by statistical tests). Note that this difference between the SFG~1 and SFG~2 goes against what would be expected if the variations observed were due to an aperture effect--the same effect should have been observed in both samples. Moreover, in both samples, we observe a comparable increase in bulge masses and absolute magnitudes (cf. Figure~\ref{fig:26} and Figure~\ref{fig:27} in appendix A). Because these two observations are clearly independent of the aperture, this result suggests the masses of the galaxies increase at higher redshift. Taken at face value, therefore, our observations are comparable to those made by \citet{pilyugin11}, suggesting that the most massive galaxies formed first at high redshift.

\begin{figure}
\epsscale{1.0} \plotone{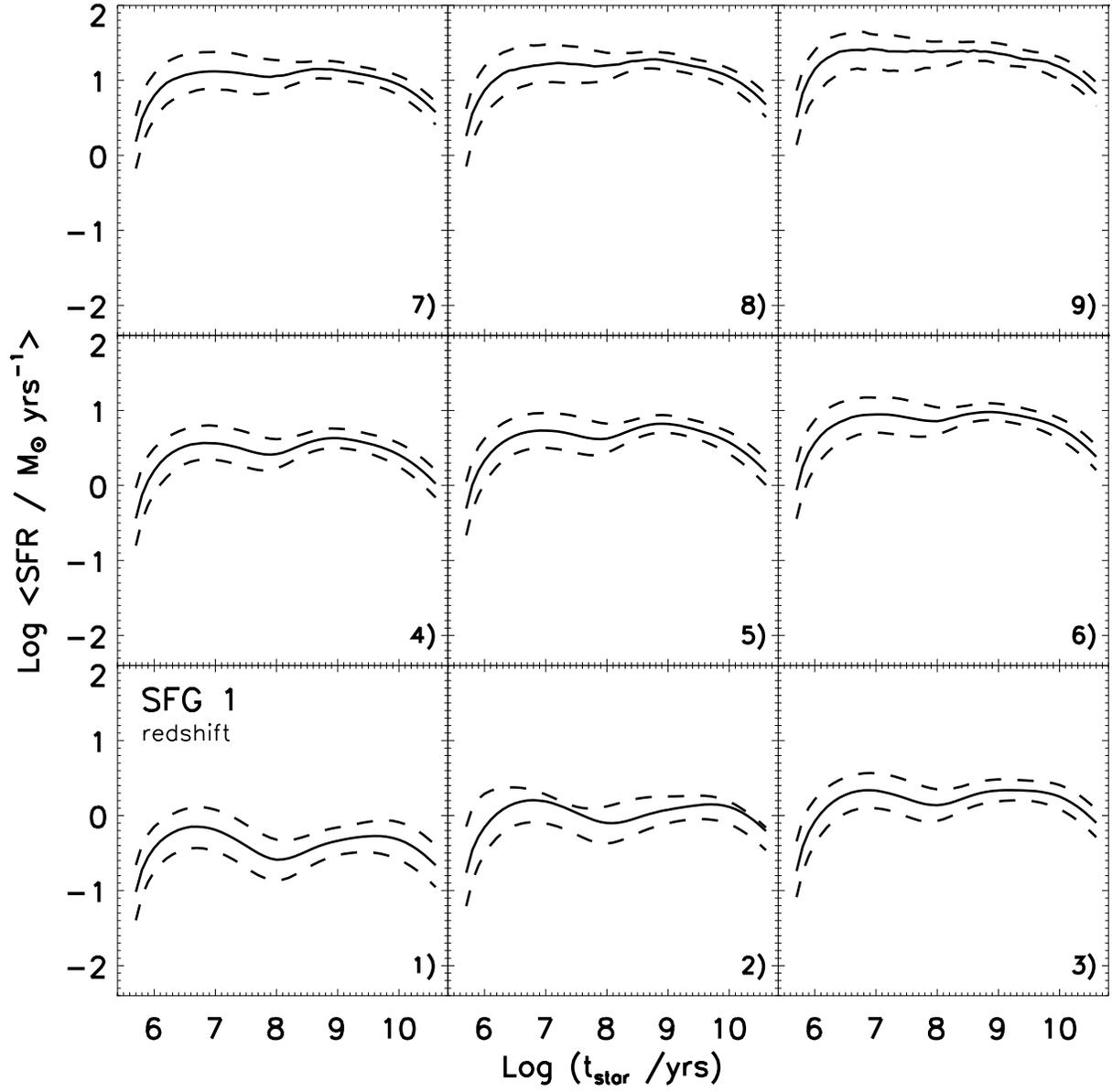} 
\caption{SFH evolution with redshift in the SFG~1. The redshift of the galaxies increases from 1) to 9).} 
\label{fig:19}
\end{figure}

\begin{figure}
\epsscale{1.0} \plotone{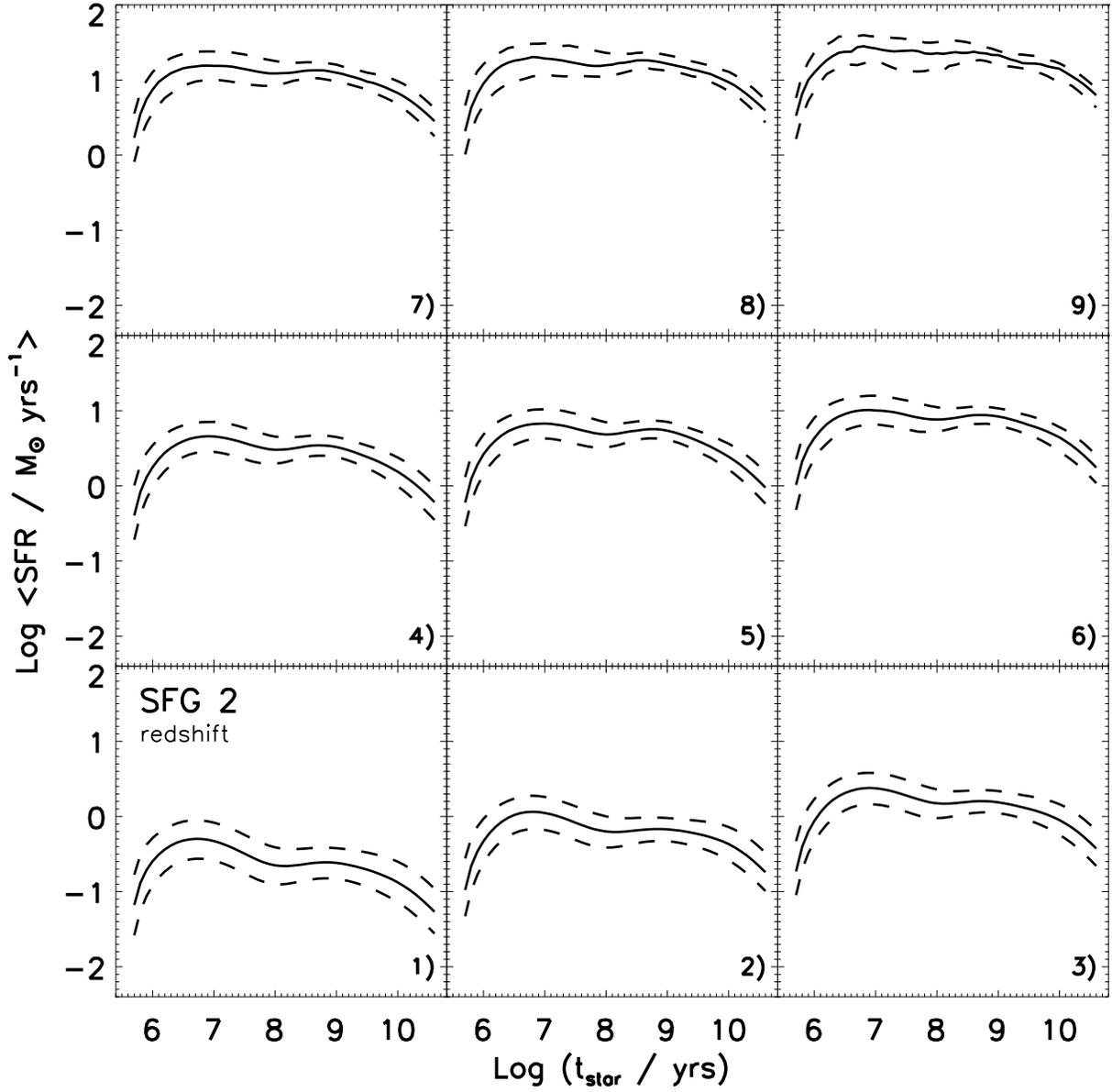}
 \caption{Same as Figure~\ref{fig:19} for the SFG~2.} 
\label{fig:20}
\end{figure}

However, we do not agree with the assumption made by \citet{pilyugin11} that the difference in abundance implies a decrease of SFR at high redshift. For example, in Figure~\ref{fig:17} and Figure~\ref{fig:18} we observe an increase of late-type spirals and of younger stellar populations at higher redshift (cf. Figure~\ref{fig:26} and Figure~\ref{fig:27} in appendix A). These trends seems somewhat in contradiction with the other observations described above. Indeed, how can the bulges increase in mass at high redshifts while the number of late-type spirals and number of young stellar population also increase? The most plausible explanation is that, in fact, the SFR in both samples increases at higher redshift. Confirming this interpretation we show in Figure~\ref{fig:19} for the SFG~1 and Figure~\ref{fig:20} for the SFG~2 the variation of SFH with redshift. For the SFG~1, all the galaxies at any redshift show a SFH consistent with constant SFR. There is consequently no evidence of a decrease of star formation with the redshift. Similarly, all the SFG~2 at any redshift show a starburst like nature (the peak at young age is higher than the peak at old age). This difference is consistent with their excess of nitrogen: at any redshift the SFG~2 are experiencing more intense bursts than the SFG~1 and consequently show more clearly the effect of starburst winds. However, and this is the most significant result yielded by STARLIGHT, in both samples we also observe a general increase of the SFR with the redshift. This is fully consistent with downsizing, showing that the more massive galaxies formed first at higher redshifts.

Note that according to our analysis the downsizing phenomenon can be observed at almost any small increment of reshift. This was unexpected and somewhat surprising. Can we be wrong in our interpretation? That is, can this be, instead, the product of some sort of luminosity bias, the galaxies are getting bigger and more massive because they are more luminous at high redshift? In our sample we do observe a decrease in the number of SFGs with the redshift (cf. Tables~7 and Tables~8), which, taken at face value, may look as an argument in favor of a luminosity bias. However, this is inconsistent with the study of \citet{Strauss02}, where the criteria for the SDSS spectroscopic survey are presented and discussed, and the authors conclude: ``The uniformity and {\bf completeness} of the galaxy sample make it ideal for studies of large-scale structure and the characteristics of the galaxy population in the local universe.'' (We emphasized the completeness.)

In fact, the explanation for the ``missing'' galaxies is simple. Our sample is indeed biased, but in another way, which is that all the galaxies are SFGs, with a S/N in line ratios $ > 3$. We verified that there are no missing galaxies in the SDSS survey at high redshift, simply they are not SFGs and do not satisfy our S/N criterion.  In particular, we have found an important number of narrow line emission galaxies with weak or missing emission lines that complete the sample. All these galaxies have earlier morphological types than the SFGs and are found in clusters and groups (the majority are consistent with low luminosity AGNs). In other words, the SDSS survey shows that the activity type of galaxies changes at higher redshift, including less and less SFGs, which is consistent with downsizing. 

Within the hierarchical biased galaxy formation paradigm, assuming the starburst phenomenon is related with galaxy formation, the SFGs, which are all spirals, are typical of the field, which means they formed relatively recently from low density fluctuations. At this level of density fluctuations the differences observed between the SFG~1 and SFG~2 must be related to very local variations, that is, on small scales of mass density, those leading to the formation of massive bulges. The downsizing phenomenon emphasizes that the redshift is really cosmological in nature. Rather than a variation of distance, it really describes a variation of state of the universe, namely here its density.

\section{Conclusion}\label{sec:5}

In \citet{Coz97}, it was reported that UV bright starburst galaxies are deficient in oxygen compared to normal late-type spirals. In \citet{Coz99} it was also reported that the same sample of starburst galaxies show a possible excess in nitrogen abundance. In the present study, using a much larger (122751) and generally selected sample of SFGs (that is, the only restriction was that they do not show the presence of an AGN), we find exactly the same phenomena, suggesting that this is a common trait of star forming galaxies, and not a peculiar characteristics of starburst galaxies.

\begin{figure}
\epsscale{1.0} \plotone{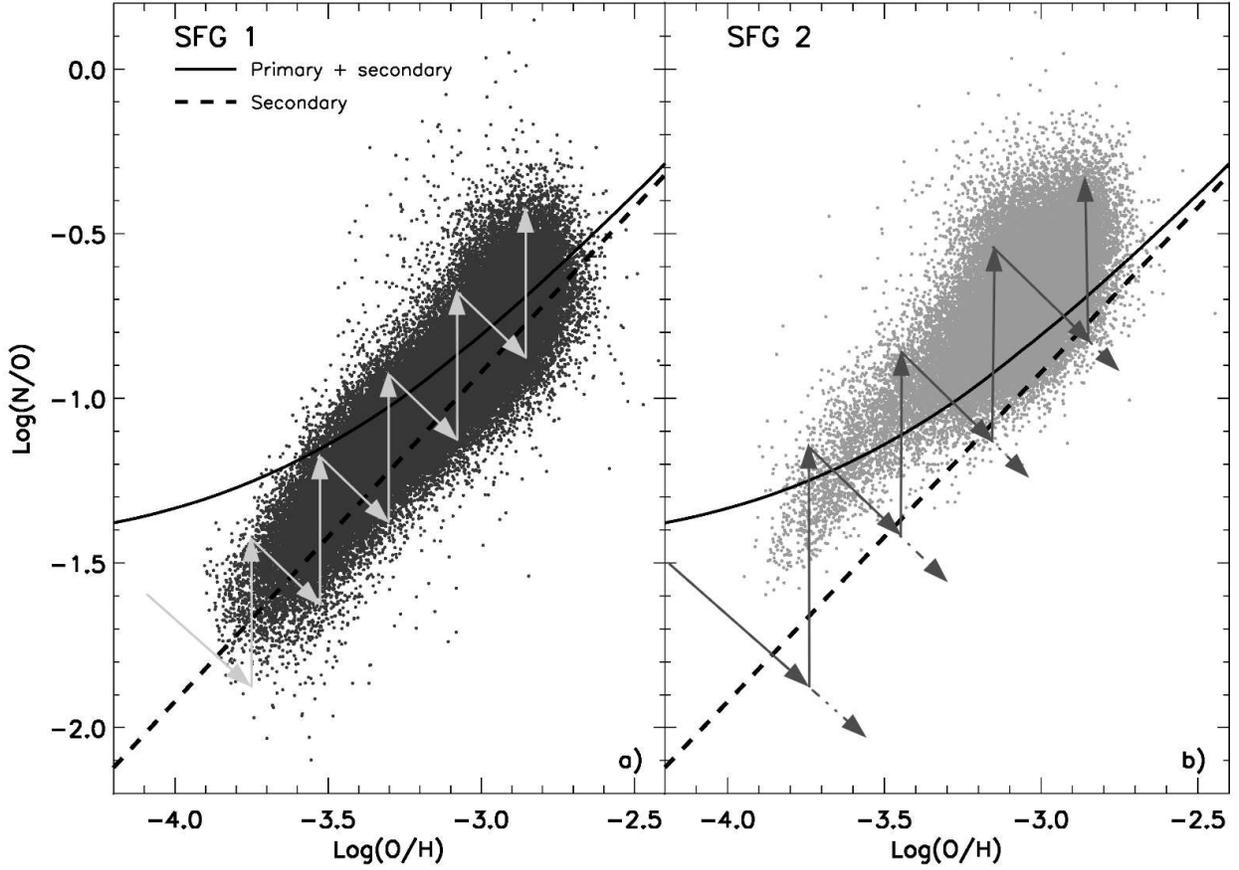}
 \caption{Illustration of the sequence of bursts model with starburst winds as main component. The star formation in the SFG~1 and SFG~2 can be reproduced by a sequence of bursts of star formation happening during their formation. Each burst causes the metallicity to increase and the ratio N/O to decrease. After $10^8$ yrs the intermediate mass stars formed during one burst begin to eject their nitrogen in the ISM, and the ratio N/O increases as the metallicity stay constant. The difference between the SFG~1 and SFG~2 is in the intensity of the bursts: in the SFG~2 more intense bursts, related to the formation of massive bulges, produce effective starburst winds, and the galaxies lose some of their oxygen. The number of intermediate stars being higher, the production of nitrogen is higher, and coupled with the lost in oxygen, a relative excess of nitrogen is observed for the same metallicity.} 
\label{fig:21}
\end{figure}

Our analysis of the SFH using STARLIGHT, in good agreement with Starburst 99 models, suggests that the depletion in oxygen and the relative excess in nitrogen are most probably due to the effect of intense starburst winds \citep{Heckman90,lehnert96,Heckman03,tremonti04,Strickland09} happening during a prolonged sequence of bursts of star formation. According to our analysis, all the SFGs form their stars through a sequence of bursts and what produces the chemical differences is a variation in intensity of the bursts. The SFGs experiencing the more intense bursts (with a median intensity above a particular threshold), or those that have experienced more intense bursts in the past, all show the effect of starburst winds: they are deficient in oxygen and relatively rich in nitrogen (N/O). We illustrate our model in Figure~\ref{fig:21}. 

Contrary to expectation, however, we find no evidence that the efficiency of the starburst wind decreases with the mass of the galaxy. Instead, our data suggests the intensity of the bursts grows with the mass, and with it the depletion of oxygen and the excess of nitrogen. To explain this observation, we propose that the abundance characteristics of these galaxies took shape during their formation process, when their gravitational wells were still forming, and consequently less massive and susceptible to decrease the efficiency of starburst winds. Consistent with this interpretation we have shown that the galaxies with an excess of nitrogen are the more massive, have bigger bulges and have an earlier morphological type than those without an excess. We have also found evidence consistent with downsizing, suggesting that the most massive galaxies formed first at higher redshift.

Considering the generality of our analysis we conclude that the formation process of the SFGs is an open process, the galaxies loosing mass and energies to their environment.

\acknowledgments

J.P. Torres-Papaqui acknowledges PROMEP for support grant 103.5-10-4684, and DAIP-Ugto (0065/11). The authors acknowledge an anonymous referee for important comments and suggestions that help us improve the quality of our article, by testing more thoroughly our hypothesis and reinforcing our arguments. Funding for the SDSS and SDSS-II is provided by the Alfred P. Sloan Foundation. The SDSS is managed by the Astrophysical Research Consortium (ARC) for the Participating Institutions. The Participating Institutions are: the American Museum of Natural History, Astrophysical Institute Potsdam, University of Basel, University of Cambridge (Cambridge University), Case Western Reserve University, the University of Chicago, the Fermi National Accelerator Laboratory (Fermilab), the Institute for Advanced Study, the Japan Participation Group, the Johns Hopkins University, the Joint Institute for Nuclear Astrophysics, the Kavli Institute for Particle Astrophysics and Cosmology, the Korean Scientist Group, the Los Alamos National Laboratory, the Max-Planck-Institute for Astronomy (MPIA), the Max-Planck-Institute for Astrophysics (MPA), the New Mexico State University, the Ohio State University, the University of Pittsburgh, University of Portsmouth, Princeton University, the United States Naval Observatory, and the University of Washington.

\appendix

\section{Results of statistical tests}

The statistical framework of the tests we used is based on a new parametric ANOVA model introduced by \citet{Hothorn08} and developed for the R software\footnote{http://CRAN.R-project.org} by \citet{Herberich10}. The max-t test does the simultaneous pairwise comparisons of means under control of the family-wise error rate, which is the probability of falsely rejecting the initial hypothesis (i.e., finding a significant difference among the means of any two groups in the data set even though there is actually no difference present). The test takes into account possible heteroscedasticity and the unequal sizes of the groups. 

We present the results of the tests under the form of simultaneous confidence intervals for all pairwise comparisons of group means. Confidence intervals including zero indicate no statistically significant differences. Confidence intervals near zero suggests some level of similarity.  The farther from zero the more significant the differences. Note the smallness of some of the confidence intervals. This is due to the very large number of data used in each bin, which makes the tests extremely significant.

\begin{figure}
\epsscale{1.0} \plotone{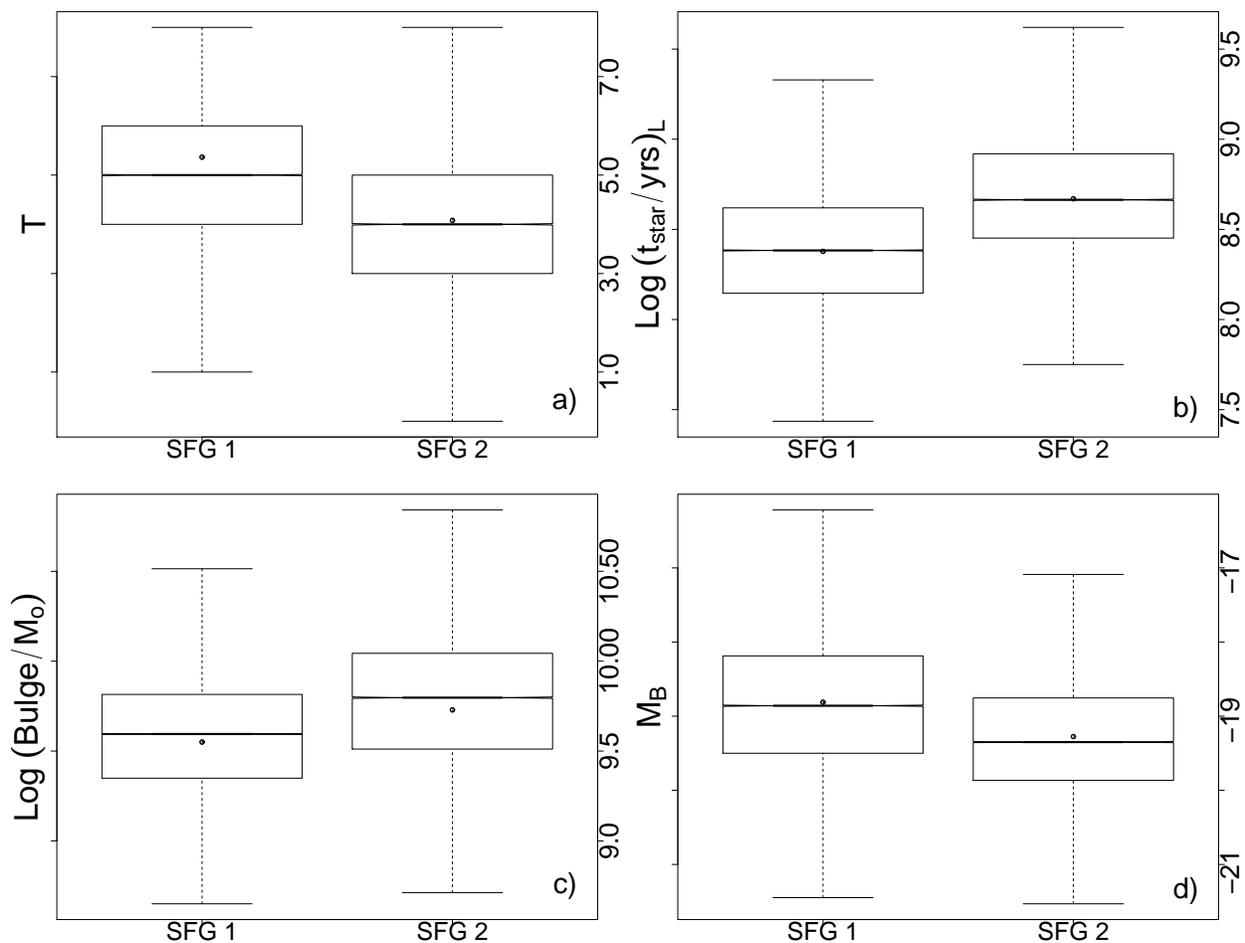} 
\caption{Box-whisker plots comparing the characteristics of the galaxies in the SFG~1 and SFG~2 samples: a) morphologies, b) mean stellar population ages, c) bulge masses, and c) absolute magnitudes in B. The points correspond to the means. Notches (barely visible due to the large size of the samples) that intersect suggest no significant difference in median.}
\label{fig:8}
\end{figure}

\begin{figure}
\epsscale{0.8} \plotone{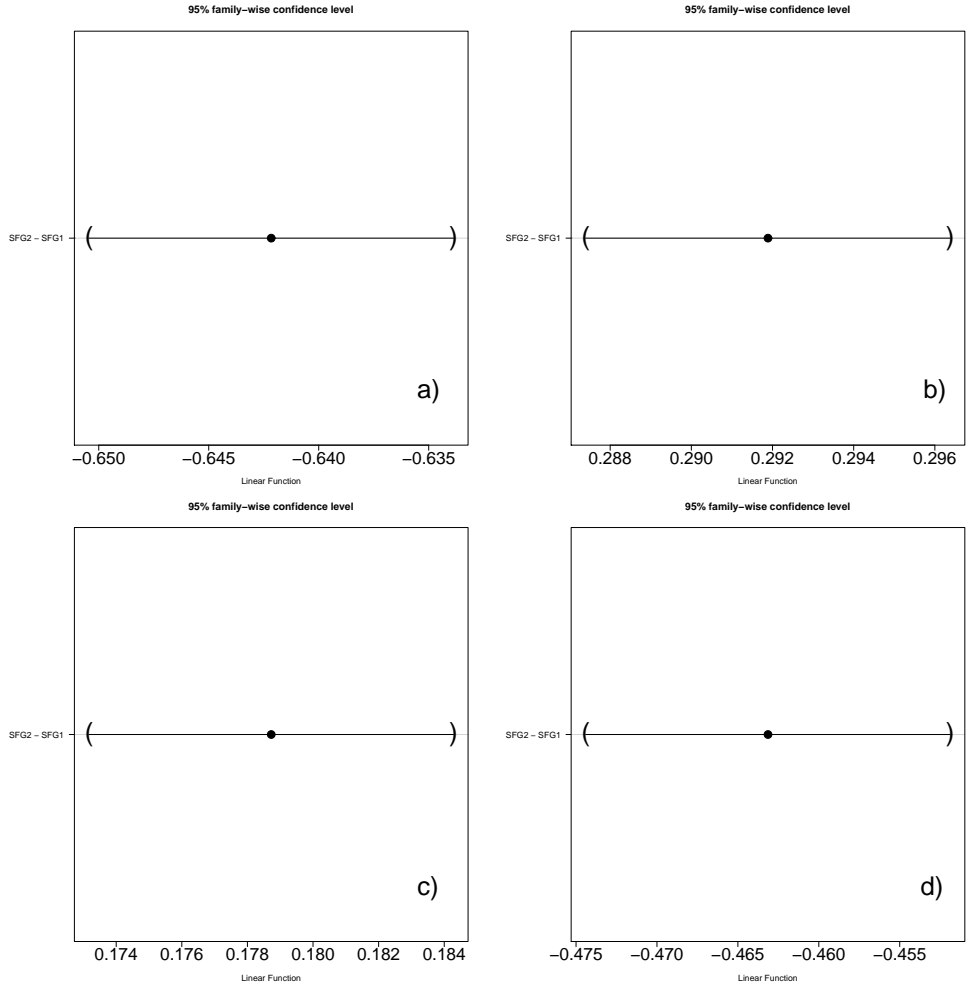}
 \caption{The confidence intervals associated with Figure~\ref{fig:8} in Section~\ref{sec:3.1}): a) morphologies; b) stellar population ages; c) bulge masses; d)  luminosity in B.} 
\label{fig:22}
\end{figure}

In Figure~\ref{fig:22} we show the confidence intervals associated with Figure~\ref{fig:8} (Section~\ref{sec:3.1}). The SFG~2 galaxies, a) have more early-type, b) have older stellar populations, c) have more massive bulges, and d) are more luminous in B than the SFG~1. 

\begin{figure}
\epsscale{1.0} \plotone{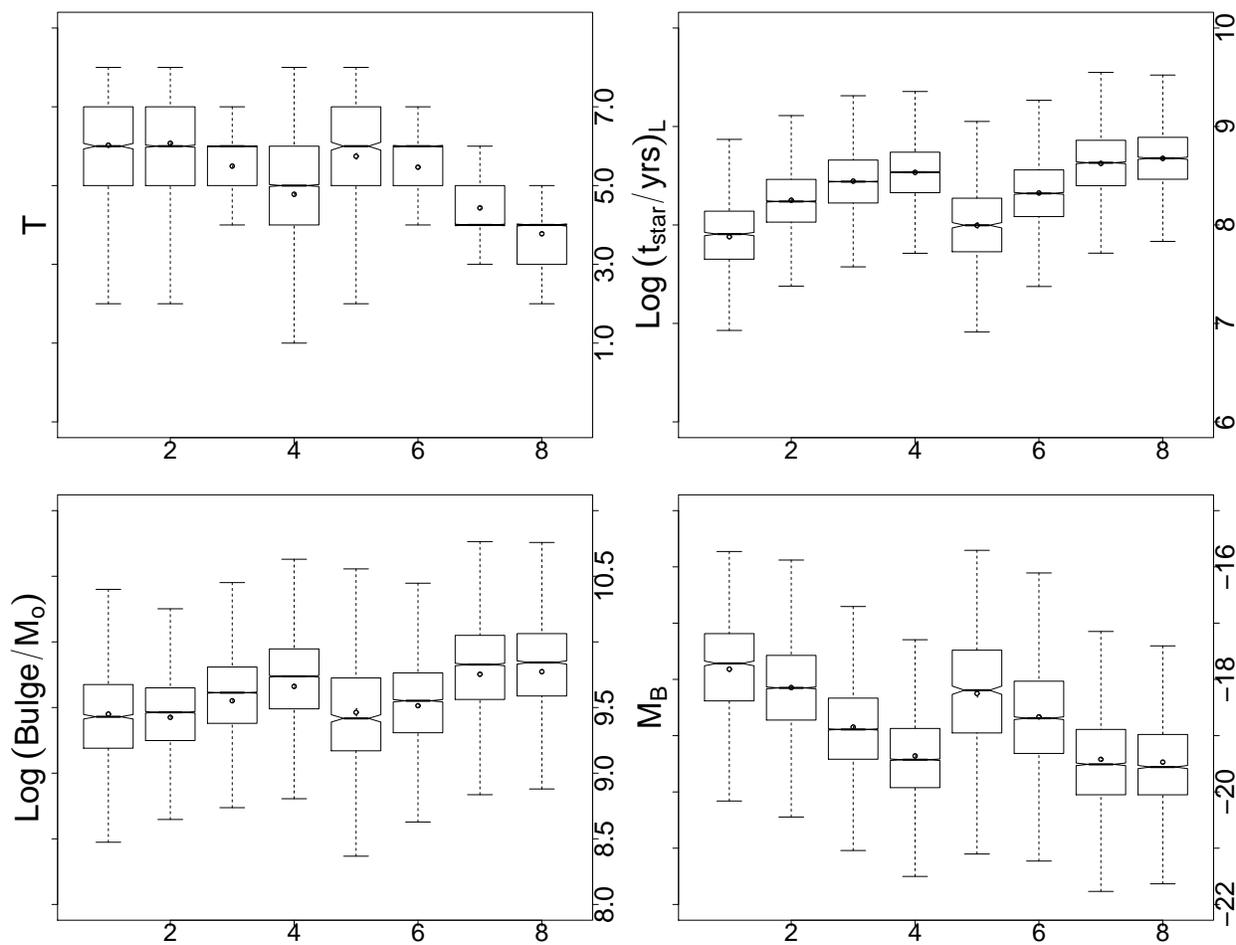}
\caption{Box-whisker plots for the 8 SFG~1 subsamples, as defined in Figure~\ref{fig:9}a. We compare: a) the morphologies, T, b) the mean stellar ages, t$_{star}$, c) the bulge masses, d) the absolute magnitudes in B. The diamond points with error bars correspond to the means. Notches that intersect suggest no significant difference in median.}
\label{fig:11}
\end{figure}

\begin{figure}
\epsscale{0.8} \plotone{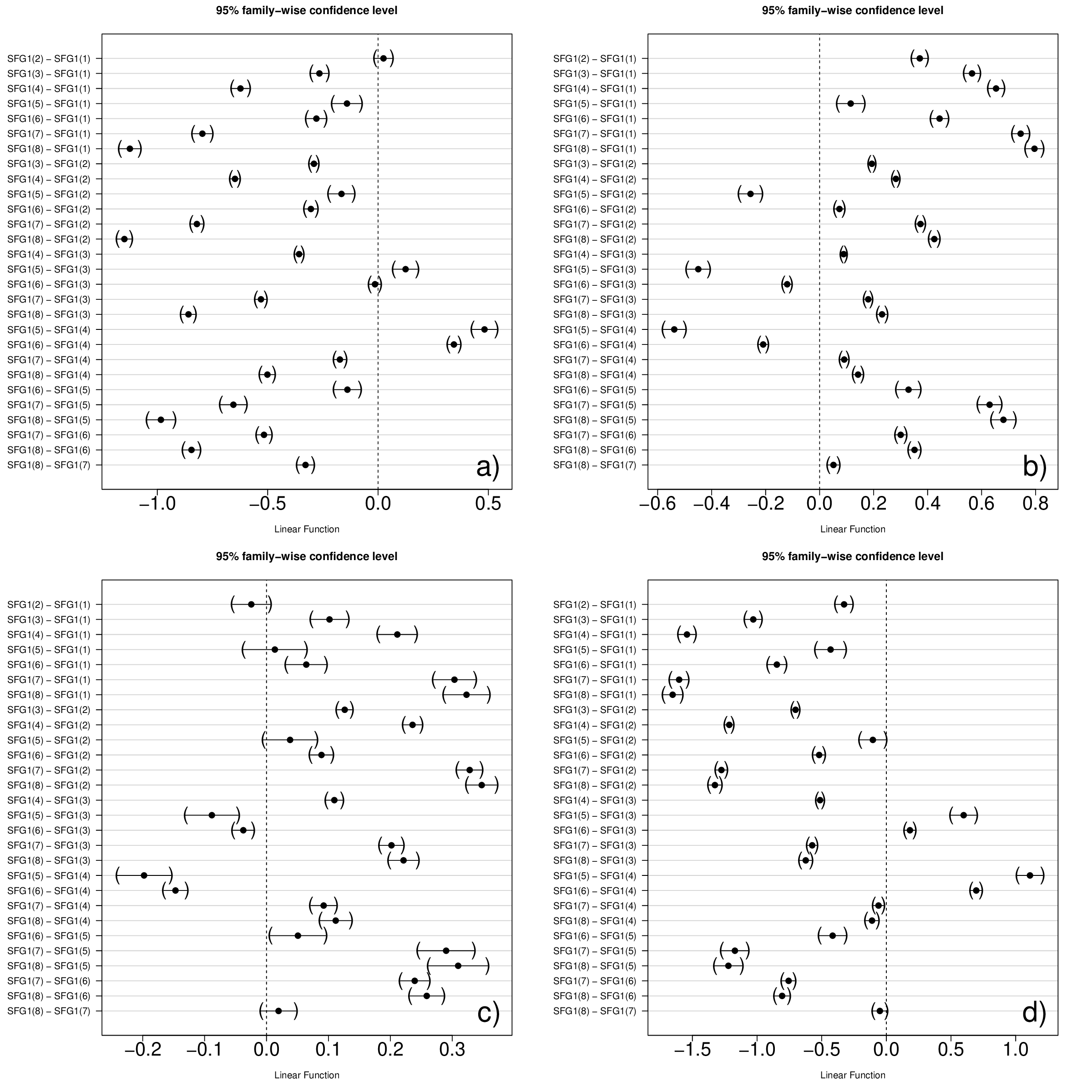}
 \caption{The confidence intervals associated with Figure~\ref{fig:11} in Section~\ref{sec:3.2}: a) morphologies; b) stellar population ages; c) bulge masses; d)  total mass.} 
\label{fig:23}
\end{figure}

In Figure~\ref{fig:23} we present the confidence intervals associated with Figure~\ref{fig:11} (Section~\ref{sec:3.2}) where we compare in the SFG~1 the variations of physical parameters in 8 subsamples of metallicities: in a) the morphologies are significantly different except in pairs of bins (1,2) and (6,3). The general trend is for an increase of metallicity in early-type galaxies; in b) all the subsamples have different stellar population ages, the general trend being an increase of age with metallicity; in c) the variation in bulge mass is less obvious (many bins show no difference) but the trend is clear the mass increasing with the metallicity; in d)  the increase in total mass with metallicity is more obvious. 

\begin{figure}
\epsscale{1.0} \plotone{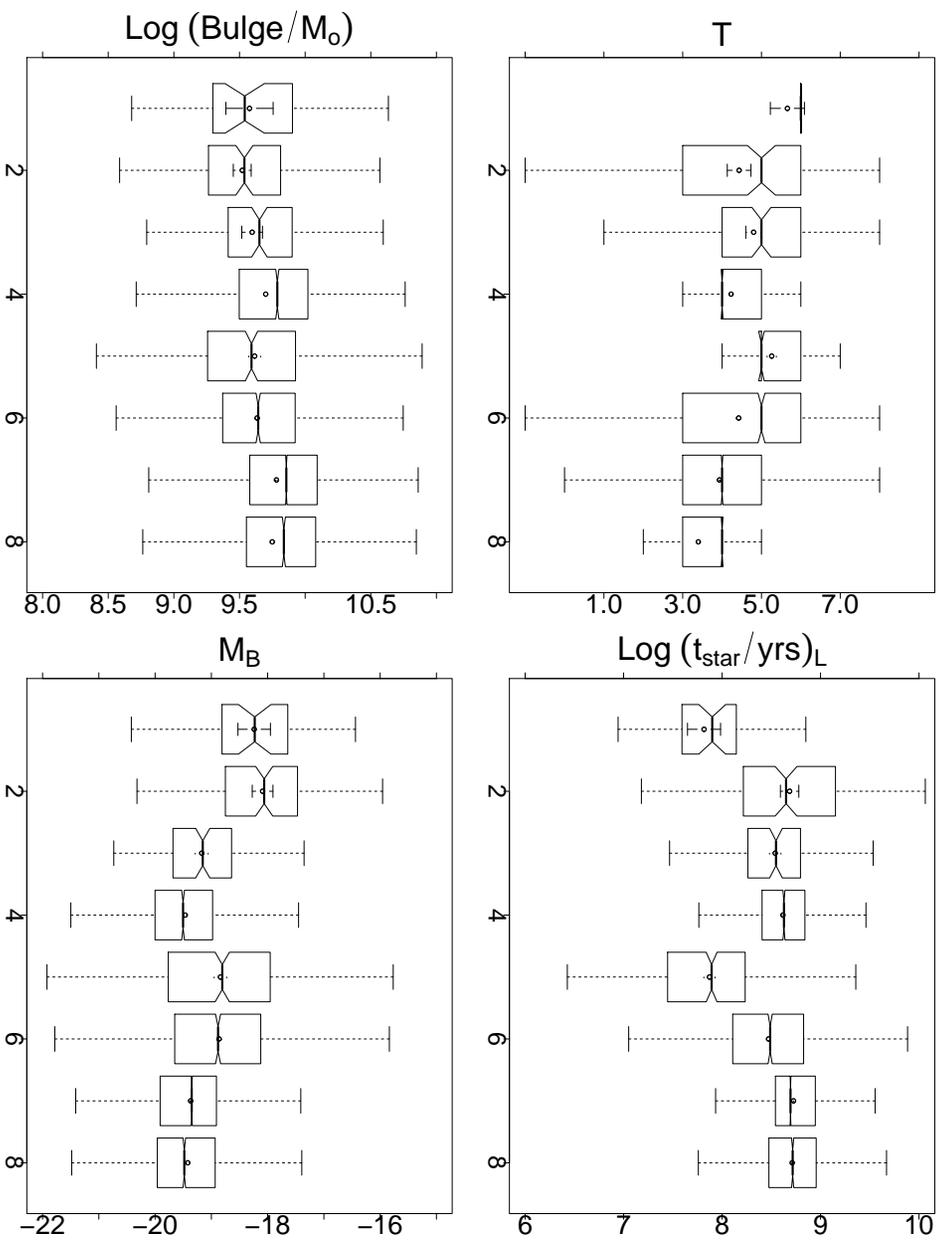} 
\caption{Same as in Figure~\ref{fig:11} for the 8 SFG~2 subsamples as defined in Figure~\ref{fig:9}b.}\label{fig:12}
\end{figure}

\begin{figure}
\epsscale{0.8} \plotone{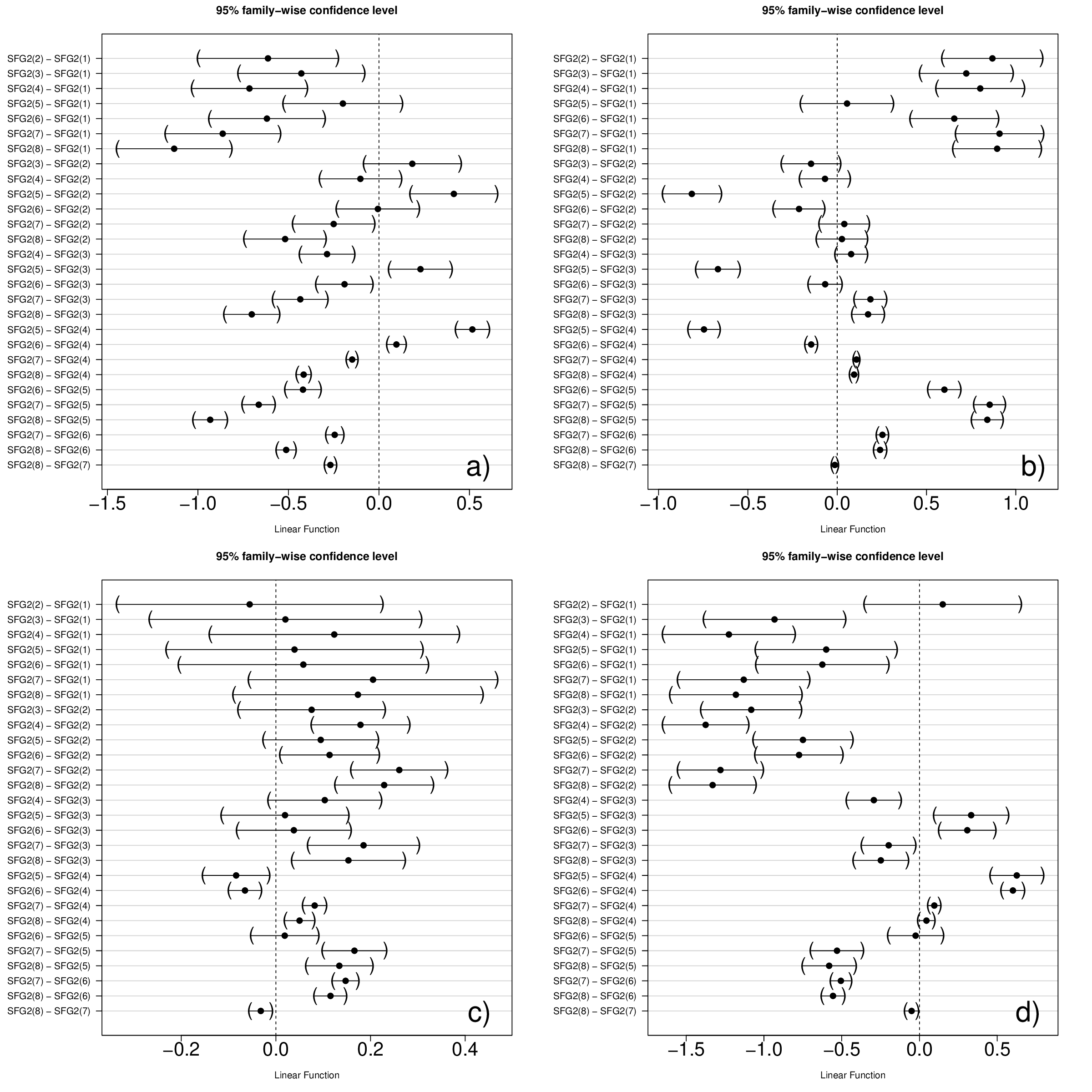}
 \caption{The confidence intervals associated with Figure~\ref{fig:12} in Section~\ref{sec:3.2}: a) morphologies; b) stellar population ages; c) bulge masses; d)  total mass. } 
\label{fig:24}
\end{figure}

In Figure~\ref{fig:24} we present the confidence intervals associated with Figure~\ref{fig:12} (Section~\ref{sec:3.2}) where we compare in the SFG~2 the variations of physical parameters in 8 subsamples of metallicities. The same trends are observed than in the SFG~1. However, the variations are more subtile, which suggests that the parameters are somewhat independent of the metallicity; that is the SFG~2 are mostly early type galaxies (a) with massive bulges (c) dominated by old stellar populations, with a mass (d) increasing with the metallicity. 

\begin{figure}
\epsscale{1.0} \plotone{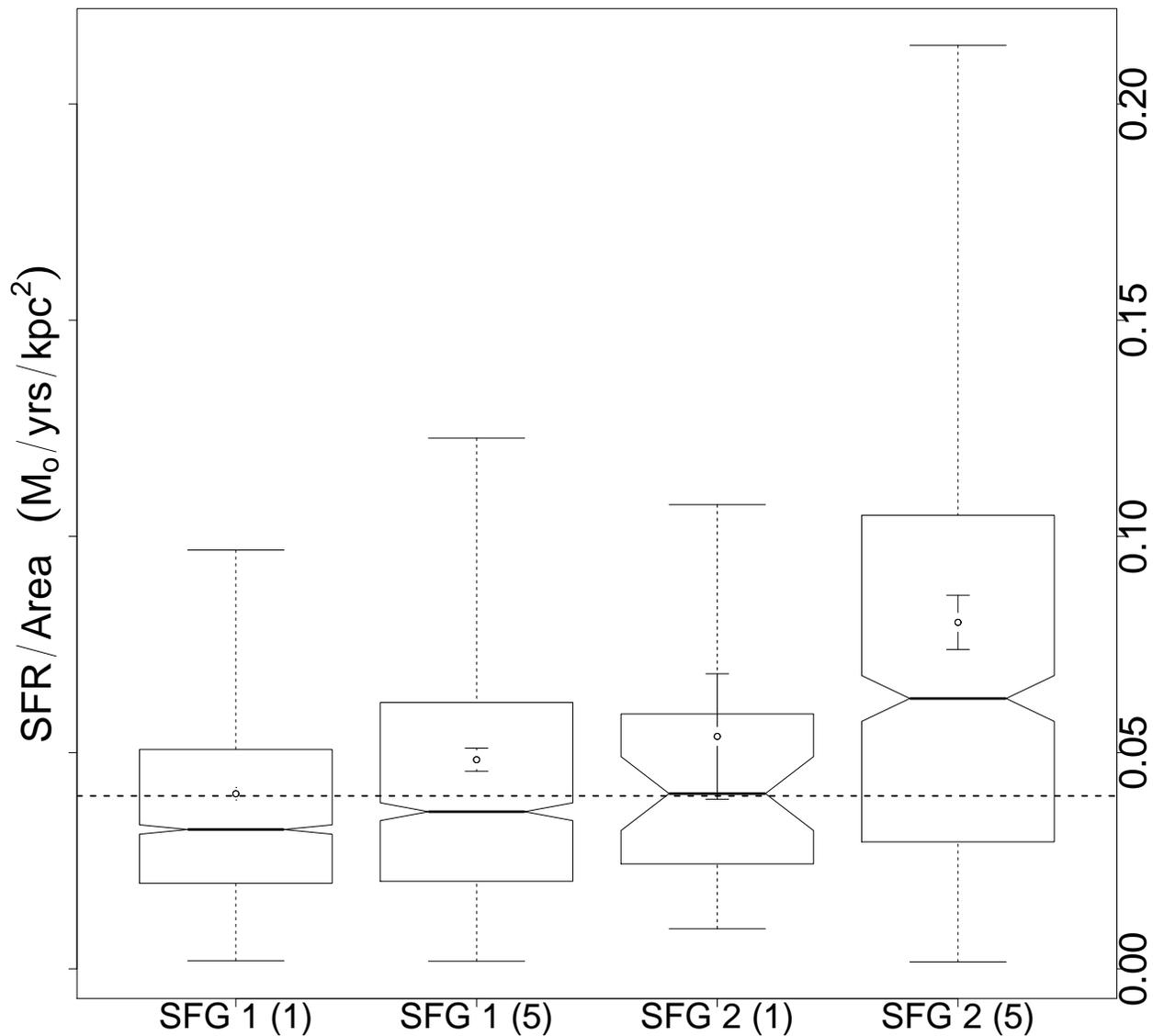} 
\caption{Box-whisker plots for the burst intensities (in unit of M$_{\odot}\ \rm{yrs}^{-1} \rm{kpc}^{-2}$) for the galaxies in both samples showing the higher recent SFR (subsamples 1 and 5). The means with estimated errors are also shown, as well as the threshold adopted by \citet{Strickland09}.}
\label{fig:16}
\end{figure}

\begin{figure}
\epsscale{0.8} \plotone{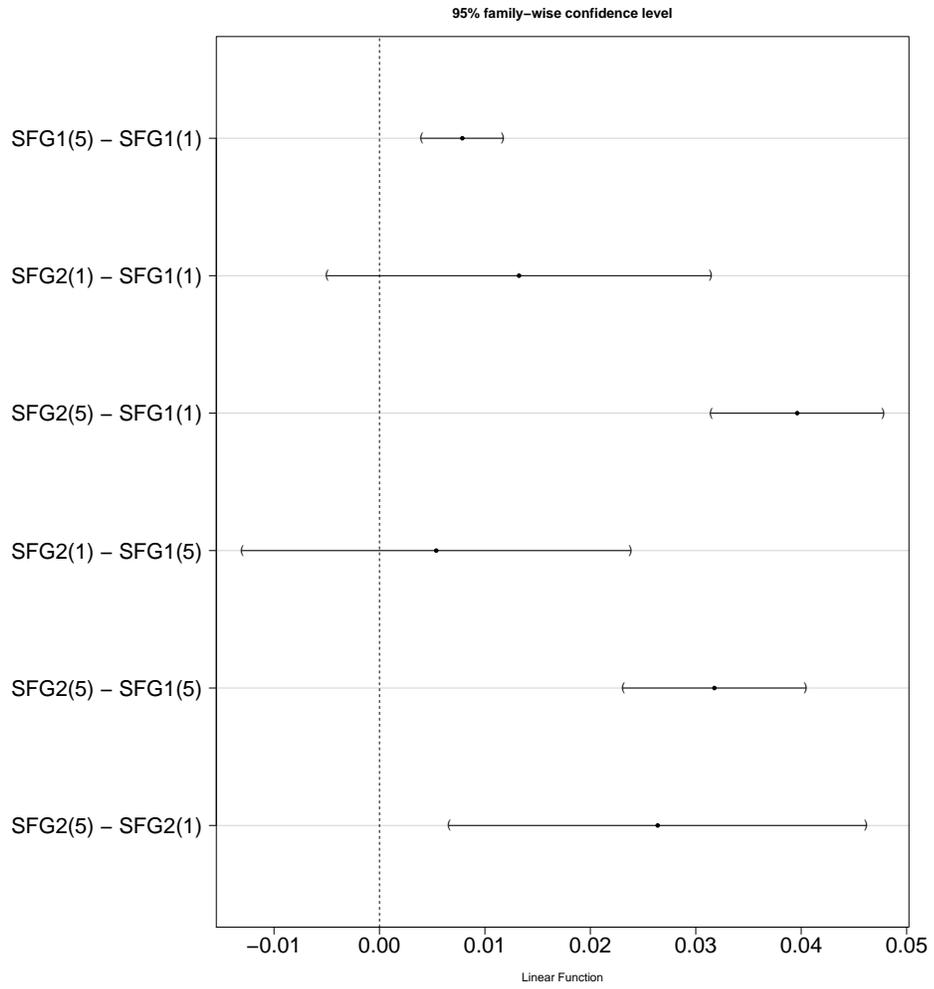}
 \caption{The confidence intervals associated with Figure~\ref{fig:16} in Section~\ref{sec:4}.} 
\label{fig:25}
\end{figure}

In Figure~\ref{fig:25} we present the confidence intervals associated with Figure~\ref{fig:16} (Section~\ref{sec:4}) where we compare the intensity of the bursts in the SFG~1 and SFG~2 in the two first bins in metallicities (1 and 5). The SFG~2 in bin 5 are clearly experiencing more intense bursts than the galaxies in all the other bins. However, also the SFG~1 in bin 5 (which are richer in nitrogen) are experiencing more intense bursts than the galaxies in bin 1 (poorer in nitrogen). In fact, the intensity of star formation in this subsample is comparable to what is observed in the SFG~2 in bin 1 (where the nitrogen abundance of the SFG~2  is comparable to that in the SFG~1)

\begin{figure}
\epsscale{1.0} \plotone{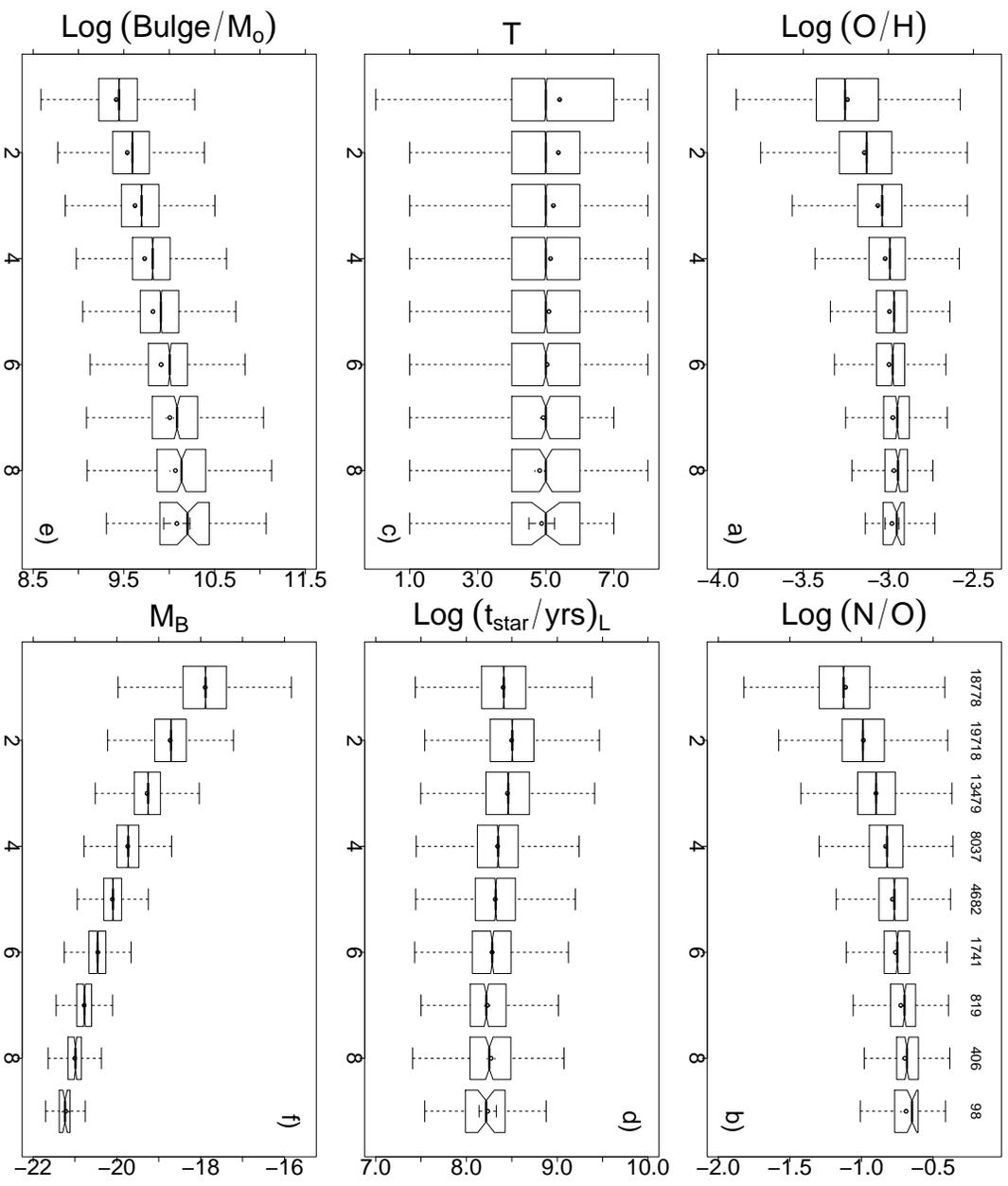} 
\caption{Box-whisker plots for the 9 SFG~1 subsamples with increasing redshifts. The redshift increases from 1 to 9.}
\label{fig:17}
\end{figure}

\begin{figure}
\epsscale{1.0} \plotone{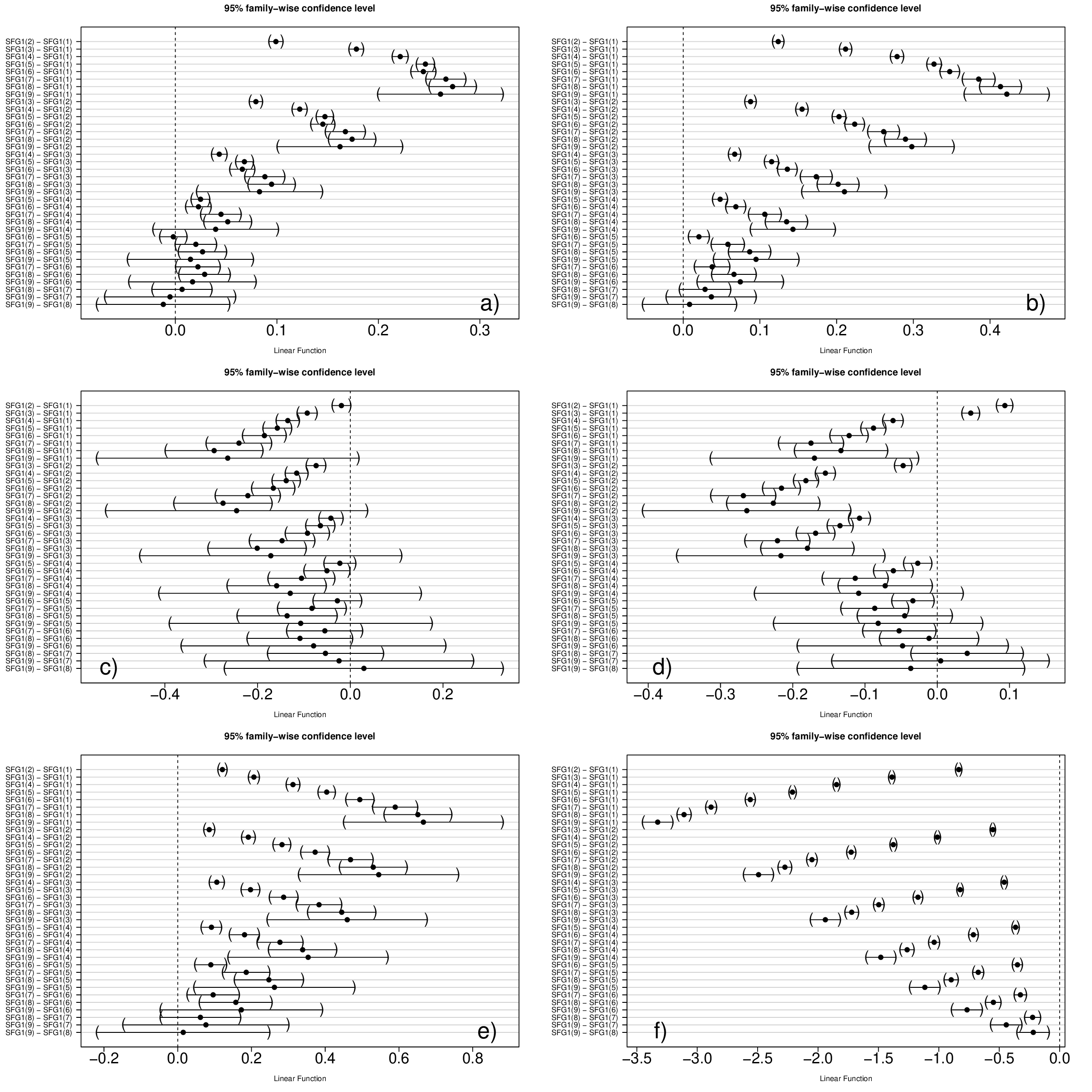}
 \caption{The confidence intervals associated with Figure~\ref{fig:17} in Section~\ref{sec:4}: a) metallicity b) nitrogen abundance; c) morphologies; d) stellar population ages; e) bulge masses; f)  total mass. } 
\label{fig:26}
\end{figure}

 In Figure~\ref{fig:26} we present the confidence intervals associated with Figure~\ref{fig:17} (Section~\ref{sec:4}) where we compare in the SFG~1 the variations of physical parameters in 9 subsamples increasing in redshift. The general trend observed in a) and b) is an increase of metallicity and nitrogen at high redshift. There are fewer variations in morphological type (c)  and stellar population (d), the trends being toward early-type but younger ages at high redshift. There is however a significant variation of bulge mass (e) and total mass (f), both increasing with the redshift. 

\begin{figure}
\epsscale{1.0} \plotone{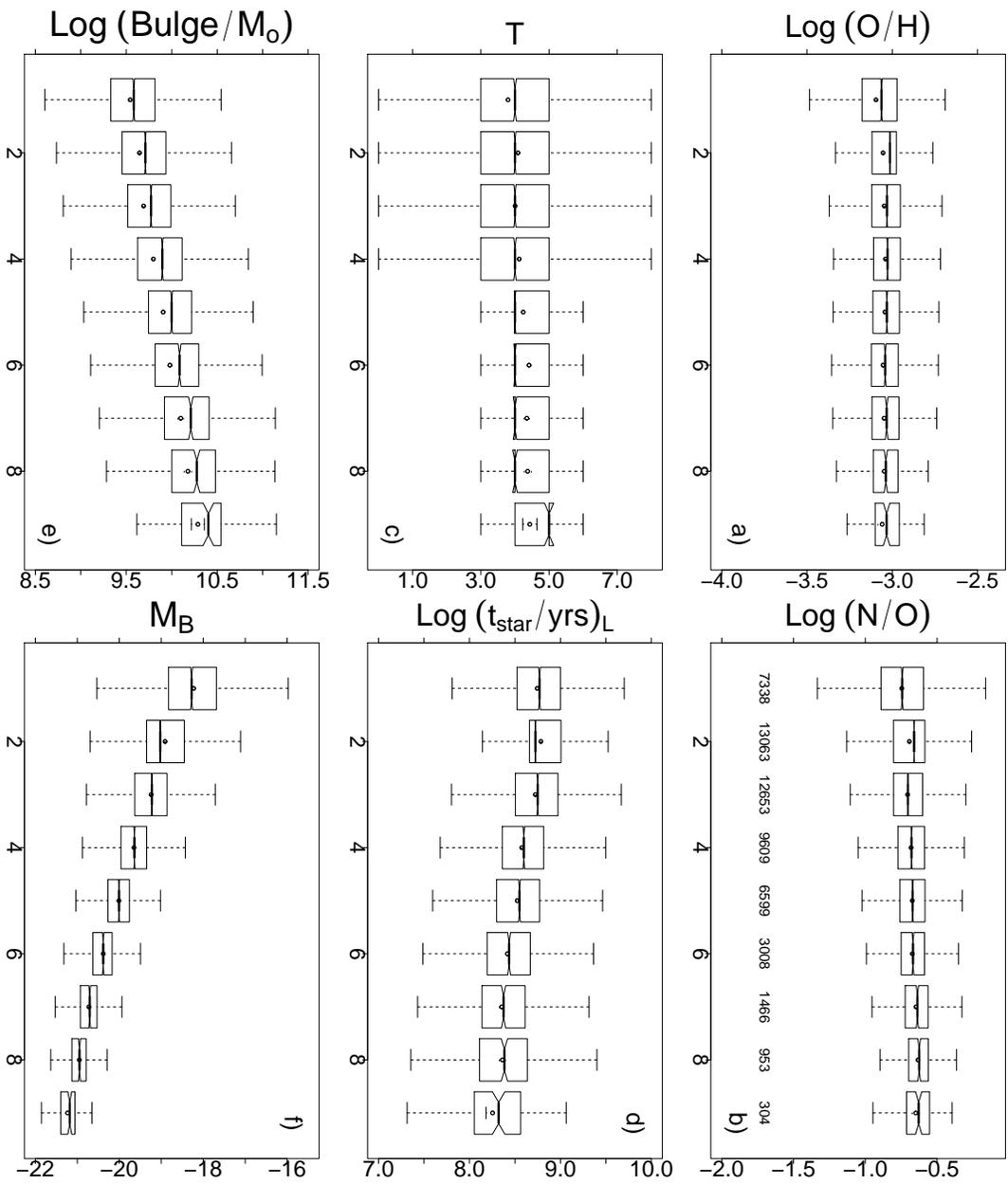} 
\caption{Same as in Figure~\ref{fig:17} for the 9 SFG~2 subsamples.} 
\label{fig:18}
\end{figure}

\begin{figure}
\epsscale{1.0} \plotone{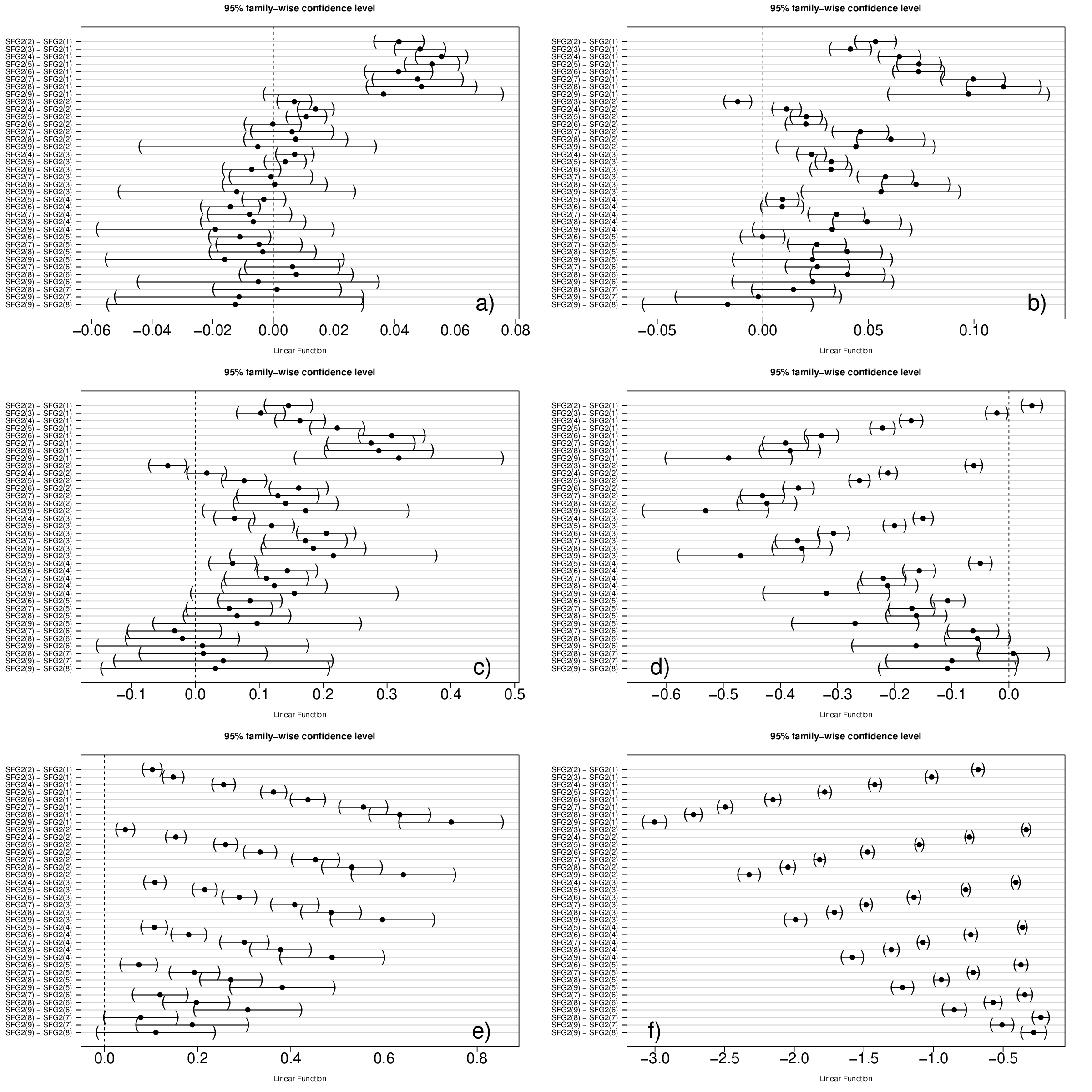}
 \caption{The confidence intervals associated with Figure~\ref{fig:18} in Section~\ref{sec:4}: a) metallicity b) nitrogen abundance; c) morphologies; d) stellar population ages; e) bulge masses; f)  total mass.} 
\label{fig:27}
\end{figure}

In Figure~\ref{fig:27} we present the confidence intervals associated with Figure~\ref{fig:18} (Section~\ref{sec:4}) where we compare in the SFG~2 the variations of physical parameters in 9 subsamples increasing in redshift. Contrary to the SFG~1 we find almost no variation of metallicity in a) and nitrogen abundance in b) with the redshift. There are very few variations in morphological type (c) and stellar populations (d), the trends being toward late-type (the contrary than in the SFG~1) and younger ages at high redshift. However, like in the SFG~1 there is a significant variation of bulge mass (e) and total mass (f), both increasing with the redshift.

\clearpage

\clearpage

\clearpage

\clearpage

\clearpage

\clearpage

\clearpage

\clearpage

\clearpage

\clearpage

\clearpage

\clearpage

\clearpage

\clearpage

\clearpage

\clearpage

\clearpage

\clearpage

\clearpage

\clearpage

\clearpage

\clearpage

\clearpage

\clearpage

\clearpage

\clearpage

\clearpage

\begin{table}
\caption{Slopes and correlation parameters for $\log$ L(H$\alpha$) vs. $\log$ L$_{\mathrm{FC}}\,\lambda 4800$.}
\begin{tabular}{lccc}
\hline
Sample     & slope      & Spearman  &  Pearson  \\
\hline
SFG 1     & $1.10\pm 0.05$ &  0.883 &  0.891 \\
SFG 2     & $1.14\pm 0.03$ &  0.932 &  0.932 \\
SFG 3     & $1.11\pm 0.05$ &  0.877 &  0.886 \\
SFG 4     & $1.13\pm 0.03$ &  0.931 &  0.931 \\
Seyfert 2 & $1.48\pm 0.03$ &  0.810 &  0.829 \\
LINER     & $1.45\pm 0.03$ &  0.739 &  0.773 \\
\hline
\end{tabular}
\\
Notes: the probability of correlation by chance is practically 0 for all the tests. 
\label{table1}
\end{table}

\clearpage

\begin{table*}
{\scriptsize
\caption{Correspondence between Hubble morphology and morphological indices, T}
\begin{tabular}{lccccccccccccc}
\hline
Hubble &  E      & E/S0     & S0 & S0/Sa & Sa  & Sab    & Sb  & Sbc   & Sc  & Scd & Sd  & Sdm  & Im \\
\hline
de Vaucouleurs & -6 to -4 & -3 to -1 & 0  &   $\dots$    & 1.0 &  $\dots$  & 3.0 &  $\dots$  & 5.0 & $\dots$ & 7.0 & 8.0 to 9.0 & 10 to 11 \\
This work & -5     & -2       & 0  & 0.5   & 1.0 &  2.0   & 3.0 & 4.0   & 5.0 & 6.0 & 7.0 & 9.0 & 10.0 \\
\hline
\end{tabular}
\label{table2}
}
\end{table*}

\clearpage

\begin{table}
\caption{Pearson ($r$) and Spearman ($r_{}$) correlation tests with chance probabilities ($P$) for the mass-metallicity and mass-nitrogen excess relations.}
\begin{tabular}{lcccc}
\hline
Pairs  & $r$ & $P(r)$ & $r_s$ & $P(r_s)$\\
\hline
M$_B$ vs [O/H] (SFG~1)  & -0.4091 & 0.0450 & -0.4369 & 0.0445 \\
M$_B$ vs [O/H] (SFG~2)  & -0.1376 & 0.4893 & -0.1390 & 0.5143 \\
M$_B$ vs $\Delta$(N/O) (SFG~1) & -0.2016 & 0.3652 & -0.2160 & 0.3865 \\
M$_B$ vs $\Delta$(N/O) (SFG~2) & -0.0638 & 0.5964 & -0.0842 & 0.5643 \\
\hline
\label{table3}
\end{tabular}
\end{table}

\clearpage

\begin{table}
{\footnotesize
\caption{Physical characteristics of SFG~1 as divided by metallicity intervals}
\begin{tabular}{lccccccc}
\hline
Bin  & Number  &    Range in  &   Range in  & T  & Log$\langle \frac{t_{star}}{yrs} \rangle_L$ & Log($\frac{M_{bulge}}{M_\odot}$)   & M$_B$ \\
id.    & of gal. &  $\log$(O/H) & $\log$(N/O) &    &                           &  &       \\
\hline
\ 1  &  1759 & $-3.85<x\leq-3.55$ & $-1.85<y\leq-1.40$ & 6.0 & 7.88 & 9.42 & -17.81 \\
\ 2  & 11375 & $-3.55<x\leq-3.25$ & $-1.60<y\leq-1.15$ & 6.0 & 8.25 & 9.44 & -18.14 \\
\ 3  & 25561 & $-3.25<x\leq-2.95$ & $-1.25<y\leq-0.80$ & 6.0 & 8.44 & 9.55 & -18.84 \\
\ 4  & 13697 & $-2.95<x\leq-2.65$ & $-1.05<y\leq-0.60$ & 5.0 & 8.53 & 9.66 & -19.36 \\
\ 5  &   861 & $-3.85<x\leq-3.55$ & $-1.40<y\leq-0.95$ & 6.0 & 7.99 & 9.46 & -18.24 \\
\ 6  &  5640 & $-3.55<x\leq-3.25$ & $-1.15<y\leq-0.75$ & 6.0 & 8.32 & 9.51 & -18.66 \\
\ 7  &  5739 & $-3.25<x\leq-2.95$ & $-0.80<y\leq-0.35$ & 4.0 & 8.62 & 9.75 & -19.42 \\
\ 8  &  3126 & $-2.95<x\leq-2.65$ & $-0.60<y\leq-0.15$ & 4.0 & 8.68 & 9.77 & -19.47 \\
\hline
\label{table4}
\end{tabular}}
\end{table}

\clearpage

\begin{table}
{\footnotesize
\caption{Physical characteristics of SFG~2 as divided by metallicity intervals}
\begin{tabular}{lccccccc}
\hline
Bin  & Number  &    Range in &   Range in  & T  & Log$\langle \frac{t_{star}}{yrs} \rangle_L$ & Log($\frac{M_{bulge}}{M_\odot}$)   & M$_B$ \\
id.   & of gal. &  $\log$(O/H) & $\log$(N/O) &    &    &  &       \\
\hline
\ 1  &    41 & $-3.85<x\leq-3.55$ & $-1.85<y\leq-1.40$ & 6.0 & 7.82 & 9.52 & -18.23 \\
\ 2  &   207 & $-3.55<x\leq-3.25$ & $-1.60<y\leq-1.15$ & 5.0 & 8.68 & 9.57 & -18.08 \\
\ 3  &  9214 & $-3.25<x\leq-2.95$ & $-1.25<y\leq-0.80$ & 5.0 & 8.53 & 9.59 & -19.17 \\
\ 4  &  5544 & $-2.95<x\leq-2.65$ & $-1.05<y\leq-0.60$ & 4.0 & 8.62 & 9.69 & -19.46 \\
\ 5  &   590 & $-3.85<x\leq-3.55$ & $-1.40<y\leq-0.95$ & 5.0 & 7.87 & 9.61 & -18.83 \\
\ 6  &  3327 & $-3.55<x\leq-3.25$ & $-1.15<y\leq-0.75$ & 5.0 & 8.47 & 9.63 & -18.86 \\
\ 7  & 30201 & $-3.25<x\leq-2.95$ & $-0.80<y\leq-0.35$ & 4.0 & 8.71 & 9.75 & -19.36 \\
\ 8  &  5863 & $-2.95<x\leq-2.65$ & $-0.60<y\leq-0.15$ & 4.0 & 8.72 & 9.78 & -19.42 \\
\hline \label{table5}
\end{tabular}}
\end{table}

\clearpage

\begin{table}
{\scriptsize \caption{Intensity of star formation and absolute magnitude of  SFGs showing evidence of recent bursts}
\begin{tabular}{lcccccccccccc}
\hline
Sample & \multicolumn{6}{c}{SFR/Area} & \multicolumn{6}{c}{M$_B$} \\
\hline
 & Min & 1Q & Median & Mean & 3Q & Max & Min & 1Q & Median & Mean & 3Q & Max \\
SFG~1 (1) & 0.002 & 0.020 & 0.032 & 0.040 & 0.051 & 0.351 & -22.79 & -18.39 & -17.72 & -17.81 & -17.19 & -16.02 \\
SFG~1 (5) & 0.002 & 0.020 & 0.036 & 0.048 & 0.062 & 0.395 & -21.51 & -18.96 & -18.21 & -18.24 & -17.50 & -16.05 \\
SFG~2 (1) & 0.009 & 0.024 & 0.040 & 0.054 & 0.059 & 0.257 & -20.42 & -18.81 & -18.23 & -18.23 & -17.64 & -16.44 \\
SFG~2 (5) & 0.002 & 0.029 & 0.062 & 0.080 & 0.105 & 0.509 & -21.92 & -19.77 & -18.85 & -18.83 & -17.98 & -16.12 \\
\hline
\label{table6}
\end{tabular} }
\end{table}

\clearpage

\begin{table}
\caption{Physical characteristics of SFG~1 as divided by redshift intervals}
\begin{tabular}{lcccccc}
\hline
Bin & Range in z & Number   &  T  & Log$\langle \frac{t_{star}}{yrs} \rangle_L$ & Log($\frac{M_{bulge}}{M_\odot}$)   & M$_B$ \\
id.   &            & of gal.  &     &                                &  &       \\
\hline
\ 1  & $0.025<z\leq0.050$ & 18778 & 5.0 & 8.40 &  9.22 & -17.86  \\
\ 2  & $0.050<z\leq0.075$ & 19718 & 5.0 & 8.49 &  9.47 & -18.68  \\
\ 3  & $0.075<z\leq0.100$ & 13479 & 5.0 & 8.43 &  9.64 & -19.21  \\
\ 4  & $0.100<z\leq0.125$ &  8037 & 5.0 & 8.31 &  9.82 & -19.67  \\
\ 5  & $0.125<z\leq0.150$ &  4682 & 5.0 & 8.26 &  9.94 & -20.00  \\
\ 6  & $0.150<z\leq0.175$ &  1741 & 5.0 & 8.21 & 10.08 & -20.34  \\
\ 7  & $0.175<z\leq0.200$ &   819 & 5.0 & 8.10 & 10.20 & -20.63  \\
\ 8  & $0.200<z\leq0.225$ &   406 & 5.0 & 8.08 & 10.26 & -20.82  \\
\ 9  & $0.225<z\leq0.250$ &    98 & 5.0 & 7.99 & 10.31 & -20.93  \\
\hline \label{table7}
\end{tabular}
\end{table}

\clearpage

\begin{table}
\caption{Physical characteristics of SFG~2 as divided by redshift intervals}
\begin{tabular}{lcccccc}
\hline
Bin & Range in z & Number   &  T  & Log$\langle \frac{t_{star}}{yrs} \rangle_L$ & Log($\frac{M_{bulge}}{M_\odot}$)   & M$_B$ \\
id.   &            & of gal.  &     &                               &  &       \\
\hline
\ 1  & $0.025<z\leq0.050$ &  7338 & 4.0 & 8.77 &  9.35 & -18.26 \\
\ 2  & $0.050<z\leq0.075$ & 13063 & 4.0 & 8.83 &  9.60 & -18.78 \\
\ 3  & $0.075<z\leq0.100$ & 12653 & 4.0 & 8.74 &  9.78 & -19.21 \\
\ 4  & $0.100<z\leq0.125$ &  9609 & 4.0 & 8.58 &  9.96 & -19.62 \\
\ 5  & $0.125<z\leq0.150$ &  6599 & 4.0 & 8.52 & 10.08 & -19.98 \\
\ 6  & $0.150<z\leq0.175$ &  3008 & 5.0 & 8.37 & 10.20 & -20.34 \\
\ 7  & $0.175<z\leq0.200$ &  1466 & 5.0 & 8.27 & 10.33 & -20.63 \\
\ 8  & $0.200<z\leq0.225$ &   953 & 5.0 & 8.20 & 10.40 & -20.83 \\
\ 9  & $0.225<z\leq0.250$ &   304 & 5.0 & 8.06 & 10.47 & -21.02 \\
\hline \label{table8}
\end{tabular}
\end{table}


\clearpage

\begin{thebibliography}{}
\bibitem[Abazajian et al.(2009)]{abazajian09} Abazajian, K. N., Adelman-McCarthy, J. K., Ag\"ueros, M. A., et al. 2009, \apjs, 182, 543
\bibitem[Asari et al.(2007)]{asari07} Asari, N. V., Cid Fernandes, R., Stasi\'nska, G., et al. 2004, \aap, 417, 751
\bibitem[Asplund et al.(2004)]{Asplund04} Asplund, M., Grevesse, N., Sauval, A. J., Allende Prieto, C. \& Kiselman, D. 2004, \aap, 417, 751
\bibitem[Baldwin, et al.(1981)]{baldwin81} Baldwin, J.A., Phillips, M.M. \& Terlevich, R. 1981, \pasp, 93, 5
\bibitem[Blanton \& Roweis(2007)]{BR07} Blanton, M. R. \& Roweis, S.  2007, \aj, 133, 734
\bibitem[Brodie \& Huchra(1991)]{brodie91} Brodie, J. P. \& Huchra, J. P. 1991, \apj, 379, 157
\bibitem[Cid Fernandes et al.(2005)]{cid05} Cid Fernandes, R., Mateus, A., Sodr\'e, L., Stasi\'nska, G. \& Gomes, J.M. 2005, \mnras, 358, 363
\bibitem[Coziol(1996)]{Coz96} Coziol, R. 1996, \aap, 309, 345
\bibitem[Coziol et al.(1997)]{Coz97} Coziol, R., Contini, T., Davoust, E. \& Consid\`ere, S. 1997, \apj, 481, L67
\bibitem[Coziol et al.(1998)]{Coz98} Coziol, R., Contini, T., Davoust, E. \& Consid\`ere, S. 1998, ASPC, 147, 219
\bibitem[Coziol et al.(1999)]{Coz99} Coziol, R., Carlos Reyes, R.E., Consid\`ere, S., Davoust, E. \& Contini, T. 1999, \aap, 345, 733
\bibitem[Coziol et al.(2001)]{CDD01} Coziol, R., Doyon, R. \& Demers, S. 2001, \mnras, 325, 1081
\bibitem[de Grijs et al.(2001)]{deGrijs01} de Grijs, R., O’Connell, R. W., \& Gallagher, J. S. 2001, AJ, 121, 768
\bibitem[de Vaucouleurs et al.(1991)]{Vaucouleurs1991} de Vaucouleurs, G., et al. 1991, Third Reference Catalogue of Bright Galaxies (Berlin: Springer)
\bibitem[Edmunds \& Pagel(1978)]{EP78} Edmunds, M. G. \& Pagel B. E. P. 1978, \mnras, 185, 77
\bibitem[Edmunds \& Pagel(1984)]{EP84} Edmunds, M. G. \& Pagel B. E. P. 1984, \mnras, 211, 507
\bibitem[Evans \& Dopita(1985)]{EvansDopita85} Evans, I. N. \& Dopita M. A. 1985, \apjs, 58, 125
\bibitem[F\"orster Schreiber et al.(2003a)]{ForsterSchreiber03a} F\"orster Schreiber, N. M., Genzel, R., Lutz, D. \& Sternberg, A. 2003a, \apj, 599, 193
\bibitem[F\"orster Schreiber et al.(2003b)]{ForsterSchreiber03b} F\"orster Schreiber, N. M., Sauvage, M., Charmandaris, V., Laurent, O., Gallais, P., Mirabel, I. F. \& Vigroux, L. 2003b, \aap, 399, 833
\bibitem[Fukugita et al.(2007)]{fukugita07} Fukugita, M., Nakamura, O., Okamura, et al. 2007, \aj, 134, 597
\bibitem[Garnett(1990)]{garnett90} Garnett, D. R. 1990 \apj, 363, 142
\bibitem[Goldader et al.(1997)] {goldader97} Goldader J. D., Joseph R. D., Doyon R. \& Sanders D. B. 1997, \apjs, 108, 449
\bibitem[Greene \& Ho(2006)]{greene06} Greene, J.E., Ho, L.C. 2006, ApJ, 641, 117
\bibitem[Heckman, Armus \& Miley(1990)]{Heckman90} Heckman, T. M., Armus, L. \& Miley, G. K. 1990, \apjs, 74, 833
\bibitem[Heckman(2003)]{Heckman03} Heckman, T. M. Rev. Mex. AA Ser. Conf., 17, 47
\bibitem[Henry et al.(2000)]{henry00} Henry, R. B. C., Edmunds, M. G. \& K\"oppen, J. 2000, \apj, 541, 660
\bibitem[Herberich et al.(2010)]{Herberich10} Herberich, E., Sikorski, J. \& Hothorn, T. 2010, PLoS ONE 5(3):e9788.doi:10.1371/journal.pone.0009788
\bibitem[Hogg et al.(2001)]{hogg01} Hogg, D. W., Finkbeiner, D. P., Schlegel, D. J. \& Gunn, J. E.  2001, \aj, 122, 2129
\bibitem[Hothorn et al.(2008)]{Hothorn08} Hothorn,T., Bretz, F. \& Westfall, P. 2008, Biom J, 50, 346
\bibitem[Izotov et al.(2006)]{izotov06} Izotov, Y. I., Stasinska, G., Meynet, G., Guseva, N. G. \& Thuan, T. X., 2006 \aap, 448, 955
\bibitem[Kauffmann et al.(2003)]{kauffmann03} Kauffmann, G., Heckman, T.M., Tremonti, C.,  et al. 2003, \mnras, 346, 1055
\bibitem[Lehnert \& Heckman(1996)]{lehnert96} Lehnert, M. D. \& Heckman, T. M. 1996, \apj, 472, 546
\bibitem[Leitherer(1999)]{leitherer99} Leitherer, C. et al. 1999, \apjs, 123, 3
\bibitem[Mayya et al.(2006)]{Mayya06} Mayya, Y. D., Bressan, A., Carrasco, L. \& Hernandez-Martinez, L. 2006, \apj, 649, 172
\bibitem[Matteucci \& Tosi(1985)]{matteucci85} Matteucci, F. \& Tosi, M. 1985, \mnras, 217, 391
\bibitem[McCall(1984)]{McCall84} McCall, M. L. 1984, \mnras, 208, 253
\bibitem[McCall et al.(1985)]{McCall85} McCall, M.L, Rybski, P.M \& Shields, G.A. 1985, \apjs, 57, 1
\bibitem[Moll\'a et al.(2006)]{molla06} Moll\'a, M., V\'{\i}lchez, J. M., Gavil\'an, M. \& D\'{\i}az, A. I. 2006, \mnras, 372, 1069
\bibitem[Nagao et al.(2006)]{nagao06} Nagao, T., Maiolino, R. \& Marconi, A.  2006, \aap, 459, 85
\bibitem[Nomura \& Kamaya(2001)]{Nomura01}Nomura H. \& Kamaya H.  2001, \aj, 121, 1024
\bibitem[Osterbrock(1989)]{osterbrock89} Osterbrock, D. E. 1989, Astrophysics of Gaseous Nebulae and Active Galactic Nuclei, University Science Books.
\bibitem[Pagel et al.(1979)]{Pagel79} Pagel, B. E. J.; Edmunds, M. G.; Blackwell, D. E.; Chun, M. S.; Smith, G. 1979, \mnras, 189, 95
\bibitem[Pier et al.(2003)]{pier03} Pier, J. R., Munn, J. A., Hindsley, R. B.,  et al. 2003, \aj, 125, 1559
\bibitem[Pilyugin et al.(2003)]{pilyugin03} Pilyugin, L. S., Thuan, T. X. \& V\'{\i}lchez, J. M.  2003, \aap, 397, 487
\bibitem[Pilyugin \& Thuan(2011)]{pilyugin11} Pilyugin, L. S. \& Thuan, T. X. 2011, \apj, 726, L23
\bibitem[Pretosian(1976)]{petrosian76}Petrosian, V. 1976. \apj, 209, L1
\bibitem[Renzini \& Voli(1981)]{renzini81} Renzini, A., Voli, M. 1981, \aap, 94, 175
\bibitem[Richer \& McCall(2008)]{richer08} Richer, M. G. \& McCall, M. L.  2008, \apj, 684, 1190
\bibitem[Shimasaku et al.(2001)]{shimasaku01} Shimasaku, K., Fukugita, M., Doi, M., et al. 2001, \aj, 122, 1238
\bibitem[Simard et al.(2011)]{simard11} Simard, L., Mendel, J. T., Patton, D. R., Ellison, S. L. \& McConnachie, A. W., 2011, ApJS, 196, 11
\bibitem[Smith \& Gallagher(2001)]{Smith01} Smith, L. J. \& Gallagher, J. S. 2001, MNRAS, 326, 1027
\bibitem[Smith et al.(2006)]{Smith06} Smith, L. J., Westmoquette, M. S., Gallagher, J. S., O’Connell, R. W., Rosario, D. J. \& de Grijs, R. 2006, \mnras, 370, 513
\bibitem[Stasinska et al.(2006)]{stasinska06} Stasinska, G., Cid Fernandes, R., Mateus, A., Sodr\'e, L. \& Asari, N. V. 2006, \mnras, 371, 972
\bibitem[Strauss et al.(2002)]{Strauss02} Strauss, M. A., Weinberg, D. H., Lupton, Robert H. and 35 coauthors 2002, AJ, 124, 1810
\bibitem[Strickland \& Heckman(2009)]{Strickland09} Strickland, D. K. \& Heckman T. M. 2009, \apj, 697, 2030
\bibitem[Thuan et al.(2010)]{thuan10} Thuan, T.~X., Pilyugin, L.~S., \& Zinchenko, I.A.  2010, \apj, 712, 1029
\bibitem[Thurston et al.(1996)]{Thurston96} Thurston, T. R., Edmunds, M. G. \& Henry, R. B. C.  1996, \mnras, 283, 990
\bibitem[Torres-Papaqui et al.(2011)]{TP11} Torres-Papaqui, J. P., Coziol, R., Ortega-Minakata, R. A. 2011, AcUni, 21, 82
\bibitem[Torres-Papaqui et al.(2012)]{TP12} Torres-Papaqui, J.P., Coziol, R., Andernach, H., et al. 2012, RMxAA, submitted
\bibitem[Tremonti et al.(2004)]{tremonti04} Tremonti, C. A., Heckman, T. M., Kauffmann, G., et al. 1992, \apj, 410, 543
\bibitem[Vacca \& Conti(1992)]{VC92} Vacca, W.D., \& Conti, P.S. 1992, \apj, 410, 543
\bibitem[van Zee et al.(1998)]{vanZee98} van Zee, L., Salzer, J. J., Haynes, M. P., O'Donoghue, A. A. \& Balonek, T. J.  1998, \aj, 116, 2805
\bibitem[Vila-Costas \& Edmunds(1993)]{VE93} Vila-Costas, M.B. \& Edmunds, M.G. 1993, \mnras, 265, 199
\bibitem[Veilleux \& Osterbrock(1987)]{veilleux87} Veilleux, S. \& Osterbrock, D.E. 1987, \apjs, 63, 295
\bibitem[Veilleux et al.(2005)]{Veilleux05} Veilleux, S., Cecil, G. \& Bland-Hawthorn, J.  2005,  ARA\&A, 43, 1056
\bibitem[Yin, Liang \& Zhang(2007)]{yin07} Yin, S. Y., Liang, Y. C. \& Zhang, B.  2007, ASPC, 373, 686
\bibitem[York et al.(2000)]{york00} York, D. G., Adelman, J., Anderson, J. E. Jr., et al. 2000, \aj, 120, 1579
\bibitem[Zaritsky et al.(1994)]{Zaritsky94} Zaritsky, D., Kennicutt, R. C. Jr. \& Huchra, J. P. 1994, \apj, 420, 87
\end{thebibliography}
\end{document}